\def\bm#1{\mbox{\boldmath{$#1$}}}
\def\comment#1{}
\newcommand{\sfrac}[2]{\raisebox{0.095ex}{\scriptsize${\frac{#1}{#2}}$}}

\documentclass[preprint,showpacs,showkeys,preprintnumbers,amsmath,amssymb]{revtex4}
\usepackage{graphicx}
\usepackage{dcolumn}
\usepackage{bm}
\usepackage{color}
\usepackage{bm}
\usepackage{subeqnarray}
\usepackage[active]{srcltx}

\def\|{|\!|}

\def\d{\textrm{d}}

\newcommand{\ba}{\begin{array}}\newcommand{\ea}{\end{array}}
\newcommand{\be}{\begin{equation}}\newcommand{\ee}{\end{equation}}
\newcommand{\bea}{\begin{eqnarray}}\newcommand{\eea}{\end{eqnarray}}
\newcommand{\brr}{\begin{array}}\newcommand{\err}{\end{array}}
\newcommand{\bit}{\begin{itemize}}\newcommand{\eit}{\end{itemize}}
\newcommand{\ben}{\begin{enumerate}}\newcommand{\een}{\end{enumerate}}

\def\bi#1{{\bm #1}}
\def\rmd{{\rm{d}}}
\def\rmi{{\rm{i}}}
\def\rme{{\rm{e}}}
\def\beq{\begin{equation}}
\def\eeq{\end{equation}}

\begin{document}

\title{Superstatistics approach to path integral for a  relativistic particle\\}

\author{Petr~Jizba}
 \email{p.jizba@fjfi.cvut.cz}
\affiliation{FNSPE, Czech Technical University in Prague,
B\v{r}ehov\'{a} 7, 115 19 Praha 1, Czech Republic\\ {\rm and} \\
ITP, Freie Universit\"{a}t Berlin, Arnimallee 14 D-14195 Berlin,
Germany\\}
\author{Hagen Kleinert}
 \email{kleinert@physik.fu-berlin.de}
\affiliation{ITP, Freie Universit\"{a}t Berlin, Arnimallee 14
D-14195 Berlin, Germany\\ {\rm and} \\
ICRANeT, Piazzale della Republica 1,\\ 10 -65122, Pescara, Italy\\
~\\}


\begin{abstract}
Superstatistics permits the calculation
of the Feynman propagator of a relativistic particle in a novel way from
a superstatistical average over non-relativistic single-particle
paths. We illustrate this for the Klein-Gordon particle in
the Feshbach-Villars representation, and for the Dirac particle in the
Schr\"{o}dinger-Dirac representation. As a byproduct we recover the
worldline representation of Klein-Gordon and Dirac propagators, and
discuss the role of the smearing distributions in fixing the
reparametrization freedom. The emergent relativity picture that follows from our
approach together with a novel representation of the Lorentz group for the Feshbach-Villars
particle are also discussed.
\end{abstract}


\maketitle

\section{Introduction}


There has been a recent upsurge of interest in the so-called
superstatistics paradigm~\cite{beck:01,beck:03,wilk:00,touchette:05,sattin:04,beck:05,vignat:05,chavanis:06,JK:08,jkh07,beck:09}.
Superstatistics is a branch of statistical
physics devoted originally to the study of non-equilibrium non-Gaussian
systems. It is characterized by superpositions
of different distribution functions which usually operate
on  vastly different time scales with a non-Gaussian distribution as an output.
Such an approach has a long tradition. There are many examples of non-linear or
non-equilibrium systems that have been
treated with such methods~\cite{Dunning-Davies:05,lavenda:91,feller66}.
The recent revival of interest in this field has been caused by recognizing the ubiquitous character of
such a statistical behavior in the nature. This has in turn
led to a systematic classification of the various compound
distributions~\cite{beck:05,beck:09}. Particularly important is the
realization that there are only three major physically relevant
universality classes of smearing distributions:
$\chi^2$-superstatistics, inverse $\chi^2$-superstatistics, and
lognormal superstatistics. These three classes arise as universal
limit statistics in majority of known superstatistical
systems~\cite{beck:05,beck:09}.

In an earlier paper~\cite{JK:08} we have shed yet another light on
the superstatistics paradigm by addressing the following question:
Assume that a conditional probability distribution $P({\bi
x}_b,t_b|{\bi x}_a,t_a)$ describing a particle  to move from the
position ${\bi x}_a$ in a $D$-dimensional euclidean space at time
$t_a$ to the position ${\bi x}_b$ at time $t_b$ satisfies the
Chapman-Kolmogorov equation for Markovian process, i.e.
\begin{eqnarray}
P({\bi x}_b,t_b|{\bi x}_a,t_a) = \int \rmd {\bi x} \ \! P({\bi x}_b,t_b|{\bi x},t)
P({\bi x},t|{\bi x}_a,t_a)\, , \;\;\;\;\;\;\;\; t_b \geq t \geq t_a\, .
\label{1.1}
\end{eqnarray}
This equation implies that
$P({\bi x}_b,t_b|{\bi x}_a,t_a)$ possesses
a path integral representation (see e.g. Ref.~\cite{PI}). If we
distinguish various distributions by a strength parameter
$v$, then
\begin{eqnarray}
P_v({\bi x}_b,t_b|{\bi x}_a,t_a)
 = \int_{{\bi x}(t_a)= {\bi
x}_a}^{{\bi x}(t_b) = {\bi x}_b} {\mathcal{D}}{\bi x} \
\!\frac{\mathcal{D} {\bi p}}{(2\pi)^D} \ \!
\exp\left\{\int_{t_a}^{t_b} \rmd \tau[\rmi {\bi p} \cdot\dot{{\bi x}} -
v H({\bi p},{\bi x}) ]\right\}\! .
\end{eqnarray}
Is it possible that also superpositions of such path integrals
\begin{eqnarray}
\bar{P}({\bi x}_b,t_b|{\bi x}_a,t_a) = \int_{0}^{\infty} \rmd v \ \!
\omega(v, t_{ba}) P_v({\bi x}_b,t_b|{\bi x}_a,t_a)
\label{1.6s}\end{eqnarray}
satisfy the Chapman-Kolmogorov equation~(\ref{1.1})? The answer is affirmative, if
the weight function $\omega (v,t)$ fulfills a certain simple
functional equation. In Ref.~\cite{JK:08} we have derived
this equation as follows. We have first defined a rescaled weight
function
\begin{equation}
w(v,t)\ \equiv\  \omega (v/t,t)/t\, ,
\label{1.6b}\end{equation}
and calculated its 
Laplace transform
\begin{eqnarray}
\tilde w(p_v,t)\ \equiv \
\int_0^{\infty} \rmd v  \ \! e^{-p_vv}w(v,t)\, .
\label{1.6c}\end{eqnarray}
The condition that the superposition (\ref{1.6s}) satisfies
(\ref{1.1}) can then be recast as
a simple
factorization property
\begin{equation}
\tilde w(p_v,t_1+t_2)
\ = \ \tilde w(p_v,t_2)
\tilde w(p_v,t_1)\, .
\label{1.6d}\end{equation}
Assuming continuity in $t$, the solution of (\ref{1.6d}) is
unique and can be written as an exponential of some
real ``Hamiltonian" $H_v(p_v)$:
\begin{equation}
\tilde w(p_v,t)\ = \
 \rme^{-tH_v(p_v)}\, ,
\label{1.6e}\end{equation}
where $H_v(p_v)$ must increase monotonically
for large $p_v$,
and must satisfy the normalization condition
$H_v(0)=0$.
The Laplace inverse of $\tilde w(p_v,t)$ yields the desired smearing
function $\omega(v,t)$. This allows for a rather large variety of
functions $\omega(v,t)$ to guarantee the Chapman-Kolmogorov equation
(\ref{1.1}). Once this is satisfied,  the smeared distribution (\ref{1.6s})
possesses a path integral representation on its own,
associated with a new Hamiltonian $\bar{H}({\bi p},{\bi
x})$.

In Ref.~\cite{JK:08} we have exploited this relationship in
the converse direction by observing that the path integral
representing of the euclidean version of the probability {\em
amplitude\/} for a {\em relativistic scalar particle\/}  to move
from a position ${\bi x}_a$ at time $t_a$ to position ${\bi x}_b$ at
time $t_b$
\begin{eqnarray}
P({\bi x}_b,t_b|{\bi x}_a,t_a)=\int_{x(t_a)= x_a}^{x(t_b) = x_b}
{\mathcal{D}}{\bi x} \frac{\mathcal{D} {\bi p}}{(2\pi)^D} \ \!
\exp\left\{\int_{t_a}^{t_b}\!\! \rmd \tau \ \!\left[\rmi {\bi p}
\cdot\dot{\bi x} - c\sqrt{{\bi p}^2 + m^2 c^2}\right]\right\},
 \label{1.4a}
\end{eqnarray}
can be considered as a superposition of non-relativistic
free-particle path integrals, namely
\begin{eqnarray}\!\!
P({\bi x}_b,t_b|{\bi x}_a,t_a)\!=\!\!\int_{0}^{\infty}\!\!\!\rmd v \
\! \omega(v,t_{ba})\!\! \int_{x(t_a)= x_a}^{x(t_b) = x_b}
\!\!\!\!{\mathcal{D}}{\bi x}  \  \! \frac{\mathcal{D} {\bi
p}}{(2\pi)^D} \ \! \exp\left\{\int_{t_a}^{t_b}\!\!\! \rmd \tau\
\![\rmi {\bi p} \cdot\dot{\bi x} - v ({\bi p}^2 c^2 + m^2
c^4)]\right\}\!.
 \label{1.4}
\end{eqnarray}
Here ${\bi p}$ and ${\bi x}$ are vectors in $D$-dimensional
euclidean space. The weight function $\omega(v,t)$ is the Weibull
distribution~\cite{Weibull51,feller66} of order $1$ (also known
as the scaled inverse $\chi^2$ distribution~\cite{Sorensen:02}). The
Weibull distribution of order $a$ is defined by
\begin{eqnarray}
\omega(v,a,t) \ = \ \frac{a \exp\left(-a^2t/4v\right)}{2\sqrt{\pi}
\sqrt{v^3/t}}, \;\;\;\;\;\; a \in \mathbb{R}^+\, , \label{weib.aa}
\end{eqnarray}
with $\omega(v,t) \equiv \omega(v,1,t)$. From the superstatistics
point of view, the relation (\ref{1.4})
belongs to the inverse $\chi^2$-superstatistics universality class.

In the literature, the representation (\ref{1.4a}) is often referred
to as the Newton-Wigner propagator~\cite{hartle:01}.
The name Klein-Gordon kernel used in Ref.~\cite{Grosche:98}
is misleading since the propagator of
the Klein-Gordon field must include also {\em
negative} energy spectrum, reflecting the existence of
the charge-conjugated solution --- antiparticle.

It is the purpose of this paper to use the superstatistics relationship
to find the Feynman propagator of the Klein-Gordon field in a novel way.
Subsequently, the same method will be applied to the Dirac field.
The superstatistics approach will allow us
to circumvent the technically involved procedure of constrained
quantization~\cite{PI,polyakov:87,Gitman:90,Sundermeyer:1982gv}
that is inherent to any theory with reparametrization invariance.
The result will be the worldline representations for the two relativistic
propagators.

For a better understanding of the upcoming result we begin
by introducing, in Section~\ref{SEc2}, the Feshbach-Villars
representation of a Klein-Gordon particle. After this we present, in
Section~\ref{SEc2b}, a derivation of the corresponding
Feynman Green function in euclidean spacetime from the superstatistics standpoint.
It will be seen that one must invoke the St\"{u}ckelberg-Feynman
interpretation of antiparticles as particles of negative energy
running in the reverse time direction in order to make sense of the
Feshbach-Villars time evolution operator. On the mathematical side,
the St\"{u}ckelberg-Feynman interpretation is necessary to ensure
that the Feshbach-Villars evolution operator forms a strongly
continuous semigroup. As a byproduct, we obtain the well-known
worldline representation of the Klein-Gordon propagator
\cite{PI,polyakov:87}. The same approach produces
also the propagator of the Dirac particle, due to a close analogy between
Feshbach-Villars diagonalisation, which brings the Hamiltonian into
a form where the positive and negative energy parts are explicitly
separated, and the Foldy-Wouthuysen transformation of Dirac's
Hamiltonian. This is demonstrated in some detail in
Section~\ref{SEc3}. In Section~\ref{SEc4}, we discuss the role of
smearing distributions in fixing the reparametrization freedom. In
particular, we show that  Weibull's distribution
parameter $a$ is closely related to the {\it einbein\/} characterizing the worldline.

In Section~\ref{SEc5a}, we briefly comment on the relation
of our superstatistical path integrals to the concept of ``emergent relativity".  Various remarks
and generalizations are proposed in the concluding
Section~\ref{SEc5}. For reader's convenience we relegate some
technical issues concerning the Feshbach-Villars representation to two
Appendices.

\section{Spinless particle in Feshbach-Villars representation~\label{SEc2}}

We start with the observation of Feshbach and Villars~\cite{Feshbach58}
that the Klein-Gordon equation for a free spinless
charged particle can be rewritten in a Schr\"{o}dinger-like form as
\begin{eqnarray}
\rmi\partial_t \Psi({\bi x},t) \ = {H}_{\rm FV} (\hat {\bi p}) \Psi({\bi x},t)\, , \label{2.1}
\end{eqnarray}
where $\hat {\bi p}=-\rmi\partial /\partial {\bi x}$.
The wave function  $\Psi({\bi x},t)$ is a two-component object
\begin{equation}
\Psi({\bi x},t) =
\left(
                                                                  \begin{array}{c}
 \phi({\bi x},t) \\
                                                                     \chi({\bi x},t) \\
                                                                   \end{array}
                                                                 \right),
\label{2.2abc}\end{equation}
and the Hamiltonian operator a $2\times2$-matrix
\begin{eqnarray}
{H}_{\rm FV} (\hat {\bi p})  =  (\sigma_3  +  \rmi\sigma_2) \frac{\hat {\bi
p}^2}{2m} \ + \ \sigma_3mc^2
\equiv
\hat {H}_{\rm FV}
. \label{2.2}
\end{eqnarray}
To see the equivalence with the Klein-Gordon equation, we rewrite
(\ref{2.1}) for the to components as
\begin{eqnarray}
&&\rmi\partial_t (\phi  +  \chi) \ = \ mc^2 (\phi  -  \chi)\, , \label{II.13a0} \\[1mm]
&&\rmi\partial_t (\phi  -  \chi) \ = \ \frac{\hat {\bi p}^2}{m} \ \!
(\phi  +  \chi) \ + \ mc^2 (\phi  +  \chi)\, , \label{II.13a}
\end{eqnarray}
from which we obtain
\begin{eqnarray}
(\square +  m^2c^2)(\phi  +  \chi) \ = \ 0 \;\;\;\ \mbox{and}
\;\;\;\ (\square + m^2c^2)(\phi  - \chi) \ = \ 0\, ,
\end{eqnarray}
showing that both $\phi$ and $\chi$ obey a Klein-Gordon equation
of mass $m$.

The physical role of the components $\phi$ and $\chi$ can be
understood by introducing the electromagnetic potential ${\bi
A}({\bi x},t)$ via the minimal substitution $\hat {\bi p}\rightarrow
\hat {\bi p}-e {\bi A}({\bi x},t)/c$, and noting that the
charge-conjugated wave function has the form~\cite{Feshbach58}
\begin{eqnarray}
\Psi_{\rm c} ({\bi x},t)\ = \ \sigma_1\Psi^*({\bi x},t) \ =
 \ \left(
                                                                   \begin{array}{c}
                                                                     \chi^* ({\bi x},t)\\
                                                                     \phi^*({\bi x},t) \\
                                                                   \end{array}
                                                                 \right).
\end{eqnarray}
Thus the two-component form of the wave function reflects the presence of particles
and antiparticles of opposite charge.

If we define the conjugate Hamiltonian operator as
\begin{eqnarray}
\hat{\bar{H}}_{\rm FV} \ \equiv \  \sigma_3 \hat {H}_{\rm
FV}^{\dagger}\sigma_3\, , \label{16.a}
\end{eqnarray}
then we see  from (\ref{2.2})  that $\hat {{H}}_{\rm FV}$
is conjugate to itself, i.e., {\it hermitian\/}
under the scalar product in $D$ spatial dimensions:
\begin{eqnarray}
(\Psi,\Psi')\ \equiv \ \int {\rm d}{\bi x} \ \! \Psi^{\dagger}({\bi
x},t)\sigma_3\Psi' ({\bi x},t), ~~~~{\rm d}{\bi x}\ \equiv \ {\rm
d}^Dx\, . \label{17a}
\end{eqnarray}

The Hamiltonian ${H}_{\rm FV}({\bi p})$ can be diagonalized via the
similarity transformation
\begin{eqnarray}
{H}_{\rm FV} ({\bi p}) \ &=& \ {U}_{{\bi p}}\left(
                                \begin{array}{cc}
                                {H}_{{\bi p}}
\\
                                  0 & -{H}_{{\bi p}}
\\
                                \end{array}
                              \right)
 {U}_{{\bi p}}^{-1}
\ = \
 {U}_{{\bi p}}\ \!\sigma_3{U}_{{\bi
p}}^{-1} {H}_{{\bi p}} \, , \label{2.3}
\end{eqnarray}
with
\begin{eqnarray}
 {H}_{{\bi p}}\equiv  c
\sqrt{{\bi
p}^2 + m^2c^2},
\label{@}\end{eqnarray}
and ${U}_{{\bi p}}$ denoting the non-unitary hermitian matrix
\begin{eqnarray}
{U}_{{\bi p}} \ &=& \ \frac{(mc^2 + {H}_{{\bi p}}) \sigma _0 + (mc^2
-
 {H}_{{\bi p}}) \ \!\sigma_1}{2\sqrt{mc^2 {H}_{{\bi p}}}}
\ =
 \  \frac{(1+ \gamma_{{\bi v}})\sigma_0 + (1- \gamma_{{\bi v}})\sigma_1}{2\sqrt{\gamma_{{\bi v}}}}
 \nonumber \\[2mm]
&=& \ \exp\!\left(\mbox{-$\frac{1}{2}$}\ \!\sigma_1\ln \gamma_{{\bi
v}} \right)
 \ = \ \exp\!\left[\mbox{$\frac{1}{2}$}\ \! \sigma_1\ \!{\rm
 arcosh}\left(\mbox{$\frac{1}{2}$}(\gamma_{{\bi v}} + 1/\gamma_{{\bi v}})\right) \right]
. \label{2.3a}
\end{eqnarray}
Here $ \sigma _0$ is Pauli's two-dimensional unit matrix,
and $ \gamma_{{\bi v}}$ the usual Lorentz factor of relativistic motion
 $\gamma_{{\bi v}} \equiv (1 -{{\bi v}}^2/c^2)^{-1/2}=
H_{{\bi p}}/mc^2$, where ${\bi v} = c^2{\bi p}/H_{\bi p}$ is the
velocity of the particle. Note that the similarity transformation
$U_{{\bi p}}$ converts the non-hermitian Hamiltonian matrix ${H}_{\rm FV}$
into the hermitian Hamiltonian matrix $ \sigma _3H_{{\bi p}}$. In this form the
positive- and negative-energy solutions are decoupled. We
also observe that if $\Psi$ is a positive-energy eigenstate of
the operator $\hat {H}_{\rm FV}$, then the associated
charge-conjugated wave function $\Psi^c$ corresponds to a
negative-energy eigenstate of $\hat {H}_{\rm FV}$, and vice versa. This
is because a positive-energy solution of momentum  ${\bi p}$ can be written as
\begin{eqnarray}
\Psi_{\bi p}({\bi x})
 \ = \ u(p) e^{i{\bi p}{\bi x}}
 \ \equiv \ U_{{\bi p}} \left(
                                        \begin{array}{c}
                                         1  \\
                                          0 \\
                                        \end{array}
                                      \right)
e^{i{\bi p}{\bi x}}, \label{20a}
\end{eqnarray}
while the charge-conjugated solution reads
\begin{eqnarray}
\Psi^c_{\bi p}({\bi x}) \ = \ \sigma_1 \Psi^*_{\bi p}({\bi x}) \ = \ \sigma_1
U_{{\bi p}} \ \! \sigma_1 \left(
                                        \begin{array}{c}
                                          0 \\
                                         1 \\
                                        \end{array}
                                      \right)
e^{-i{\bi p}{\bi x}} \ = \
 v(p) e^{-i{\bi p}{\bi x}}
\ \equiv \ U_{{\bi p}}
                                      \left(\begin{array}{c}
                                          0 \\
                                          1 \\
                                        \end{array}
                                      \right) e^{-i{\bi p}{\bi x}}, \label{20b}
\end{eqnarray}
showing that it corresponds
the negative
energies. The two-component objects
\begin{eqnarray}
\xi(+\sfrac{1}{2})=
   \left(\begin{array}{c}
                                          1 \\
                                          0 \\
                                        \end{array}
                                      \right),~~~
\;\;\;\;\;\;\;\xi(-\sfrac{1}{2})=
   \left(\begin{array}{c}
                                          0 \\
                                          1 \\
                                        \end{array}
                                      \right),~~~
\label{20c}\end{eqnarray}
play the role of ``pseudospinors"  in a charge space.

From Eqs. (\ref{17a}), (\ref{20a}), and (\ref{20b}) we see that
the wave functions can be normalized according to
\begin{equation}
 (\Psi,\Psi)\ = \ \pm1\, ,
\label{21a}\end{equation}
where the plus/minus sign corresponds to particle/antiparticle. Here
we have used the relation $U_{{\bi p}} \sigma_3 U_{{\bi p}} = \sigma_3$.

Equation (\ref{20a}) suggests that ${U}_{{\bi p}}$ may be viewed as
a boost transformation that brings a ``pseudospinor" of a spinless particle
at rest, $\xi(\sfrac{1}{2})$, to the ``pseudospinor"  $u_{{\bi p}}$ of
a particle with velocity ${\bi v}$. As usual, the Lorentz boosts are, in contrast to rotations,
non-unitary, as all finite-dimensional representations of noncompact
group transformations should be. However, as in the case of Dirac spinors, they are
pseudounitary with respect to the conjugation operation (\ref{16.a}), namely $U^{-1}_{\bi p} =
\sigma_3U^{\dag}_{\bi p} \sigma_3 \equiv \bar{U}_{\bi p}$. More
details will be provided in Appendices~A and B.


\section{Feynman Green function \label{SEc2b}}

Let us calculate the Feynman Green function ${\mathcal{G}}(x,y)$
associated with the Schr\"{o}dinger-like equation (\ref{2.1}). It is
defined by
\begin{eqnarray}
({\rmi}\partial_t - \hat {H}_{\rm FV})\ \!{\mathcal{G}}({\bi x},t;{\bi
x}',t') \ = \ {\rm i} \delta^{(D)}(\bi x-\bi x') \delta(t-t') \, .
\label{2.2aa}
\end{eqnarray}
The solution has the Fourier decomposition
\begin{eqnarray}
\mbox{\hspace{-16mm}}{\mathcal{G}}({\bi x},t;{\bi x}',t')\! &=&  {\rm
i}
 \int_{\mathbb{R}^{D+1}}
\frac{\rmd p_0}{2\pi} \frac{\rmd {\bi p} }{(2\pi)^D}
\rme^{-ip(x-x')} \left[p_0 c - (\sigma_3 + \rmi\sigma_2) \frac{{\bi
p}^2}{2m}
-\sigma_3 m c^2 \right]^{-1} \nonumber \\[2mm]
\!&=&  {\rm i}\int_{\mathbb{R}^{D+1}} \frac{\rmd^{D+1} p
}{(2\pi)^{D+1}} \ \! \frac{\rme^{-ip(x-x')}}{p^2c^2 -m^2 c^4 + \rmi
\epsilon} \ \!\left[p_0 c + (\sigma_3 + \rmi\sigma_2) \frac{{\bi
p}^2}{2m} +\sigma_3 m c^2 \right]\! . \label{2.2ab}
\end{eqnarray}
The Feynman boundary conditions are ensured by
the usual $\rmi \epsilon$-prescription with infinitesimal $ \epsilon >0$.
Equivalently, we can perform a Wick
rotation, which makes the denominator regular, and the
imaginary-time Green function $\mathcal{G}({\bi x},-it;{\bi
x}',-it') \equiv P({\bi x},t|{\bi x}',t')$
satisfies the Fokker-Planck-like equation
\begin{eqnarray}
(\partial_t +\hat  {H}_{\rm FV}) P({\bi x},t|{\bi x},t') \ = \
\delta(t-t') \delta^{(3)}(\bi x-\bi x')\, . \label{2.2aab}
\end{eqnarray}
The solution is obtained from the local matrix element of the
time evolution operator $e^{-t
\hat H_{\rm FV}}$:
\begin{eqnarray}
P({\bi x},t|{\bi x}',t')\  = \ \langle {\bi x}|\ \! \rme^{-(t-t') \hat
{H}_{\rm FV}}| {\bi x}'\rangle\, .
\end{eqnarray}
Recalling the matrix relation (\ref{2.3}), this is equal to
\begin{eqnarray}
P({\bi x},t|{\bi x}',t') \ = \
\langle {\bi x}|{U}_{\hat {\bi{p}}}\ \! {\rm e}^{-(t-t')\sigma_3
{H}_{\hat {\bi p}}}\ \! {U}_{\hat {\bi{p}}}^{-1}| {\bi x}' \rangle \, .
\label{2.3a1a}
\end{eqnarray}
Since ${U}_{\hat {\bi p}}$ and ${H}_{\hat{\bi p}}$ are diagonal in the
momentum basis  $|{\bi p}\rangle$, we use the completeness relation
\begin{eqnarray}
\int_{\mathbb{R}^D} \frac{\rmd {\bi p}}{(2\pi)^D} \ \! |{\bi
p}\rangle\langle {\bi p}| \ = \ {\openone} \;\;\;\;\;\;
\end{eqnarray}
to rewrite (\ref{2.3a1a}) as
\begin{eqnarray}
&&\mbox{\hspace{-20mm}}P({\bi x},t|{\bi x}',t') =
 \int_{\mathbb{R}^D}
 \frac{\rmd {\bi p} }{(2\pi)^{D}}\ \!
\ \! {\rme^{\rmi {\bi p}\cdot({\bi x} - {\bi x}')}}
 \ \! U_{{\bi p}} \langle{\bi p}|\rme^{-(t-t')
\sigma_3{H}_{ {\bi p}}}|{\bi p} \rangle U_{{\bi p}}^{-1}\, .
\label{2.3a1}
\end{eqnarray}
Alternatively we may rewrite (\ref{2.3a1a}) as
\begin{eqnarray}
P({\bi x},t|{\bi x}',t') \ = \
{U}_{\hat {\bi{p}}} \langle {\bi x}|\ \! {\rm e}^{-(t-t')\sigma_3
{H}_{\hat {\bi p}}}\ \!| {\bi x}' \rangle
(\stackrel{\leftarrow}{U}_{{\hat
 {\bi{p}}'}
})^{-1} \, , \label{2.3a1b}
\end{eqnarray}
where the arrow on top of the operator indicates the direction
in which the momentum operator $\hat {\bi{p}}$ acts.

Let us now express the amplitude in (\ref{2.3a1b})
as a path integral. This is not straightforward, since the
formal expression
\begin{eqnarray}
\langle{\bi x}''|\rme^{-(t-t')\sigma_3 {H}_{\hat{\bi p}}}| {\bi
x}'\rangle \ &=& \ \int_{{\bi x}(t') = {\bi x}'}^{{\bi x}(t) = {\bi
x}''}\!\! {\mathcal{D}}{\bi x} \ \!\frac{\mathcal{D}{{\bi
p}}}{(2\pi)^D} \ \!\rme^{\int_{t'}^{t}\rmd \tau\left[\rmi {\bi
p}\cdot\dot {\bi x} - c \sigma_3 \sqrt{{\bi p}^2 + m^2 c^2}\right]}
.\label{2.3b}
\end{eqnarray}
diverges for the lower components of the imaginary-time evolution
operator
\begin{eqnarray}
\rme^{-t\sigma_3 {H}_{{\bi p}}} \ = \ \left(
                                                                    \begin{array}{cc}
                                                                      \rme^{-t{H}_{{\bi p}}} & 0 \\
                                                                      0 &  \rme^{t
{H}_{{\bi p}}}\\
                                                                    \end{array}
                                                                  \right).
\end{eqnarray}
This difficulty can be circumvented by using
different superpositions of Gaussian path integrals, written
as in Eq.~(\ref{1.4}), both with the same
positive Hamiltonian, but with different
weight functions $\omega(v,t_{})$ for upper and lower components of
$\rme^{-t\sigma_3 {H}_{{\bi p}}}$. In particular, we write
\begin{eqnarray}
\langle{\bi x}''|\rme^{-t\sigma_3 {H}_{\hat {\bi p}}}| {\bi x}'\rangle \
= \ \int_{0}^{\infty} \rmd v \ \!\omega(v,t_{}) \int_{{\bi x}(0) =
{\bi x}'}^{{\bi x}(t) = {\bi x}''}\!\! {\mathcal{D}}{\bi x} \
\!\frac{\mathcal{D}{{\bi p}}}{(2\pi)^D} \ \!
 \rme^{\int_{0}^{t}\rmd \tau\left[\rmi {\bi p}\cdot\dot{\bi x} -
v ({{\bi p}^2c^2 + m^2 c^4)}\right]}\, ,\label{eq.26}
\end{eqnarray}
with a matrix weight function
\begin{eqnarray}
\omega(v,t) \ = \  \frac{1}{2\sqrt{\pi} \sqrt{{v^3}/{|t|}}} \left(
                                                                    \begin{array}{cc}
                                                                     \theta(t) \ \!\rme^{-t/4v} & 0 \\
                                                                      0 & \theta(-t)\ \!\rme^{t/4v} \\
                                                                    \end{array}
                                                                  \right). \label{2.33a}
\end{eqnarray}
Here we have invoked the Feynman-Stuckelberg causality
condition~\cite{Feynman:49,Stuckelberg:41,Stuckelberg:42}
that negative-energy solutions propagate backward in time.

Note that the amplitude (\ref{eq.26}) has the time-ordered form
\begin{eqnarray}
\langle{\bi x}''|\rme^{-t\sigma_3 {H}_{\hat {\bi p}}}| {\bi x}'\rangle \
&=& \ {\theta(t)}\ \!\frac{1 + \sigma_3}{2} \ \!
 \langle{\bi x}''|\rme^{-t
{H}_{\hat {\bi p}}}| {\bi x}'\rangle  \ + \ {\theta(-t)}\ \!\frac{1 -
\sigma_3}{2} \ \!\langle{\bi x}''|\rme^{t {H}_{\hat {\bi p}}}| {\bi
x}'\rangle \nonumber \\[2mm]
&=& \frac{1 + \mbox{sgn}(t)\sigma_3}{2} \ \! \langle{\bi
x}''|\rme^{-|t| {H}_{\hat {\bi p}}}| {\bi x}'\rangle  \nonumber \\[2mm]
&=& \frac{1}{2}\left(1  - \frac{ H_{\hat {\bi p}''}\ \!\sigma_3 }{
\partial_t}\right) \langle{\bi x}''|\rme^{-|t| {H}_{\hat {\bi p}}}| {\bi
x}'\rangle   \, . \label{3.34ab}
\end{eqnarray}

Representation (\ref{eq.26}) can be given a familiar relativistic form
by functionally integrating out ${\cal D}{\bi p}$ and changing
variables from $v$ to $ \nu \equiv 1/v$. This gives ($t>0$)
\begin{eqnarray}
\langle{\bi x}''|\rme^{-t {H}_{\hat {\bi p}}}| {\bi x}'\rangle \ = \
 \int_0^{\infty} \rmd \nu \ \! \frac{\rme^{-m^2c^4 t/\nu}\
\! \rme^{-t\nu/4}}{2\sqrt{\pi}\, \sqrt{\nu/t}} \int_{{\bi x}(0) =
{\bi x}'}^{{\bi x}(t) = {\bi x}''} \mathcal{D} {\bi x} \ \!
\rme^{-\int_{0}^{t}{\rmd \tau} \ \! \nu\,\dot{{\bi x}}^2/4c^2  }
 \, .\label{2.5}
\end{eqnarray}
%
By setting $\tau = \bar\lambda  \nu /{2m c^2}$, the right-hand side
becomes
\begin{eqnarray}
&& \langle{\bi x}''|\rme^{-t {H}_{\hat {\bi p}}}| {\bi
x}'\rangle=\int_0^{\infty} \rmd \lambda_{} \ \!
\frac{\rme^{-\lambda_{} mc^2/2}\ \! \rme^{- t_{}v/4} \ \!
\sqrt{2mc^2}
 \ \!t_{}}{2\sqrt{\pi}
\sqrt{\lambda_{}^3}} \int_{{\bi x}(0) = {\bi x}'}^{{\bi x}(\lambda)
= {\bi x}''} \mathcal{D} {\bi x} \ \! \rme^{-\int_{0}^{\lambda}{\rmd
\bar\lambda} \ \! {m} [{\bi x}'(\bar  \lambda )]^2/2 }\, .
\label{2.34}\end{eqnarray}
We can now use a trivial Gaussian path integral for an auxiliary
zeroth component $x_0( \lambda )$ associated with the path ${\bi x}(
\lambda )$:
\begin{eqnarray}
\sqrt{\frac{2\pi \lambda_{}}{m}}\ \! \int^{x_0(\lambda) =
ct}_{x_0(0) = 0} \mathcal{D} {x_0} \ \! \rme^{-\int_{0}^{\lambda}
\rmd \lambda \ \! m [x_0'(\bar\lambda)]^2/2} \ = \
\rme^{-mc^2t_{}^2/2\lambda_{}}\ =  \  \rme^{- t_{}v/4}\, ,
\end{eqnarray}
whose time derivative is
\begin{eqnarray}
- \partial_{t} \left[\sqrt{\frac{2\pi \lambda_{}}{m}}\ \!
\int^{x_0(\lambda) = ct}_{x_0(0) =0}\mathcal{D} {x_0} \ \!
\rme^{-\int_{0}^{\lambda} \rmd\bar \lambda \ \! m[ x'_0(\bar
\lambda)]^2/2} \right] \ = \ \frac{v}{4} \ \! \rme^{- t_{}v/4}\,
.\label{3.38a}
\end{eqnarray}
With (\ref{3.38a}) we can finally  express (\ref{3.34ab}) in the form
\begin{eqnarray}
\langle{\bi x}''|\rme^{-t_{}\sigma_3 {H}_{\hat {\bi p}}}| {\bi x}'\rangle
\ = \ (\sigma_3 H_{\bi p} -
\partial_{t}) \left[ \int_{0}^{\infty} {\rmd \lambda_{}} \
\rme^{-\lambda_{} mc^2/2} \ \! \int_{ x(0) = x'}^{{x}(\lambda) =
{{x}}} \mathcal{D} { x} \ \! \rme^{-\int_{0}^{ \lambda }d\bar\lambda
\ \! {m} [ {x}_{\mu}'(\bar\lambda)]^2/2 }\right]\! ,\label{2.5cc}
\end{eqnarray}
where ${x}_{\mu} = (ct,{\bi x})$, ${x}^2_{\mu} = c^2t^2+{\bi x}^2$
and $x_0''= ct$. The  path integral on the right-hand side describes a free
nonrelativistic particle in $D+1$ dimensions.

The expression in brackets has a Fourier representation
(see, e.g. Chapter 19 in Ref.~\cite{PI})
\begin{eqnarray}
\ \int_{0}^{\infty} {\rmd \lambda} \ \rme^{-\lambda
mc^2/2} \! \int_{ x(0) = x'}^{{x}(\lambda) = {{x}''}}
\mathcal{D} { x} \ \! \rme^{-\int_{0}^{\lambda}{\rmd
\bar{\lambda}} \ \! ({m}/{2}) (\rmd {x}^{\mu}/\rmd \bar{\lambda})^2 }
\ = \
\int_{\mathbb{R}^{D+1}} \frac{\rmd^{D+1} p}{(2\pi)^{D+1}} \ \!
\frac{\rme^{-\rmi p(x'' - x')}}{p^2 \ + \ m^2c^2}\,
  .\label{2.6ax}
\end{eqnarray}
Inserting this into formula (\ref{2.5cc}), and the result further into
(\ref{2.3a1b}), we find
\begin{eqnarray}\!\!\!\!\!\!
P({\bi x},t|{\bi x}',t') 
&=&  \int_{\mathbb{R}^{D+1}} \frac{\rmd^{D+1} p}{(2\pi)^{D+1}} \ \!
\rme^{-\rmi p(x - x')}  \ \!
 U_{\bi p}\left(\rmi p_0 c + \sigma_3 H_{\bi p}\right)
U^{-1}_{\bi p} \ \! \frac{1}{p^2+m^2c^2}\nonumber \\[3mm]
&=& \int_{\mathbb{R}^{D+1}} \ \!
 \frac{\rmd^{D+1} p}{(2\pi)^{D+1}} \ \!
\rme^{-\rmi p(x - x')}
 \left[\rmi p_0 c + (\sigma_3 + \rmi\sigma_2)
\frac{{\bi p}^2}{2m} + \sigma_3 m c^2 \right]\frac{1}{p^2+m^2c^2} .\label{IV42aa}
\end{eqnarray}
It is straightforward to verify that (\ref{IV42aa}) satisfies the differential equation (\ref{2.2aab}).
One could, of course, arrive at (\ref{2.6ax}) directly by comparing (\ref{2.3a1}) and (\ref{2.5cc})
with (\ref{2.2ab}).

Note that by reading Eq.~(\ref{2.6ax}) from right to left we obtain the
well-known path integral representation of the Klein-Gordon propagator $ \Delta
(x-x')$~\cite{Feynman50,PI}, also known as the
Feynman-Fock worldline representation. Normally this is derived with the help of a spurious
dynamical variable -- einbein, that
makes the path integral manifestly reparametrization invariant. Such
a gauge freedom is then treated with the usual methods of
constrained quantization~\cite{PI,polyakov:87,Gitman:90}. Our use of Weibull's distribution
brought us automatically to what is sometimes called Polyakov gauge~\cite{polyakov:87}
-- i.e., the gauge where the einbein variable is fixed to be the
velocity of light (for details see Ref.~\cite{PI}).  In Section~\ref{SEc4}
we shall see how $\omega(v,t)$ must be modified to account for a general gauge.
By going back from euclidean times to real times, we can now
recover the Green function associated with the  initial real-time
Schr\"{o}dinger equation (\ref{2.1}).


\section{Dirac particle and Foldy-Wouthuysen
transformation \label{SEc3}}

It is instructive to use the same procedure for calculating the
Green function of the Dirac particle. Here the role of Feshbach-Villars
diagonalisation matrix ${U}_{\bi p}$ in Eq.~(\ref{2.3a}) is played  by the $4\times 4$
Foldy-Wouthuysen transformation in spinor space~\cite{Foldy50}
\begin{equation}
V_{{\bi p}}\ = \ \rme^{-\rmi{S}_{\bi p}}\, ,~~~~~~
{S}_{\bi p} \ = \ -\rmi \bm{\gamma}
\cdot\frac{{\bi p}}{|{\bi p}|}\frac{\theta_{\bi p}}{2}\, ,\label{3.1aa}
\end{equation}
where $\bm{\gamma}$ denotes the three-vector of the hermitian
spatial Dirac matrices and
\begin{eqnarray}
\mbox{\hspace{-0mm}}
~
 \cos
\theta_{\bi p} = \frac{mc}{\sqrt{{\bi p}^2 + m^2c^2}}  =
\frac{1}{\gamma_{\bi v}}, ~~~~~\sin \theta _{\bi p} =  \frac{|{\bi
p}|}{\sqrt{{\bi p}^2 + m^2c^2}} =  \frac{|{\bi v}|}{c}\, .
\end{eqnarray}
The matrix $V_{{\bi p}}$ brings the hermitian Dirac Hamiltonian
\begin{eqnarray}
{H}_{\rm D}({\bi p}) \ = \ c\gamma^0\bm{\gamma}\cdot{{\bi p}} \ + \
\gamma^0 m c^2 \ = \ \left(
                                                                 \begin{array}{cc}
                                                                   m c & {{\bi p}}\cdot{\bm\sigma}  \\
                                                                   {{\bi p}}\cdot\bm{\sigma} & -mc \\
                                                                 \end{array}
                                                               \right)c\, ,
                                   \end{eqnarray}
to the diagonal form
\begin{eqnarray}
{H}_{\rm diag}({\bi p})\ &=& \ \left(
                \begin{array}{cc}
                  c \sqrt{{\bi p}^2 + m^2c^2} \  \sigma _0
 & 0 \\
                  0 & - c \sqrt{{\bi p}^2 + m^2c^2} \  \sigma _0
 \\
                \end{array}
              \right) \ = \ \gamma_0 {H}_{{\bi p}}\, ,
\end{eqnarray}
with the similarity transformation
\begin{eqnarray}
{H}_{\rm D}({\bi p})\ = \  V_{{\bi p}}\ \! {H}_{\rm diag}({\bi p})
V_{{\bi p}}^{-1}.~~~~ \label{3.FWT}\end{eqnarray}
Here $ \gamma _0$ is the Dirac matrix
\begin{eqnarray}
 \gamma _0 \ = \
\left(
        \begin{array}{cc}
                   \sigma _0 &0 \\
                                                                   0 & - \sigma _0\\
                                                                 \end{array}
                                                               \right) \ = \ \sigma_3\otimes\sigma_0\, ,
\label{3.2bb}\end{eqnarray}
composed of Pauli's two-dimensional unit matrices $ \sigma _0$ ($\otimes$ denotes a tensor product).

Note that the  matrix $ V_{{\bi p}}$ is now unitary, in contrast to the
spinless case where it was non-unitary but hermitian, since  there
the Hamiltonian was non-hermitian.

In the Dirac case, we may calculate the probability
$P({\bi x},t|{\bi x}',t')$ as a $4\times 4$ matrix
following the same strategy as in Section~\ref{SEc2b}. In particular, we write
 $\langle{\bi x}|\rme^{-(t-t')\gamma_0 {H}_{{\bi p}}}| {\bi
x}'\rangle$ by analogy with (\ref{eq.26}) as a superposition of
non-relativistic single particle path integrals
\begin{eqnarray}
\mbox{\hspace{-0mm}}\langle{\bi x}|\rme^{-t\gamma_0 {H}_{\hat {\bi p}}}|
{\bi x}'\rangle  \ = \  \int_{0}^{\infty} \rmd v \ \!\omega(v,t)
\int_{{\bi x}(t') = {\bi x}'}^{{\bi x}(t) = {\bi x}}
{\mathcal{D}}{\bi x} \ \!\frac{\mathcal{D}{{\bi p}}}{(2\pi)^D} \,
 \rme^{\int_{t'}^{t}\rmd \tau\left[\rmi {\bi p}\cdot\dot{\bi x} -
v ({{\bi p}^2c^2 + m^2 c^4)}\right]}\, , \label{53}
\end{eqnarray}
with a matrix of Weibull distributions:
\begin{eqnarray}
 \omega(v,t) \ = \  \frac{1}{2\sqrt{\pi}
\sqrt{{v^3}/{|t|}}} \left(
\begin{array}{cc}
                \theta(t) \ \!\rme^{-t/4v} \ \! \sigma_0 & 0 \\
                0 & \theta(-t)  \ \!\rme^{t/4v}\ \! \sigma_0 \\
                \end{array}
                \right).
%
%
\label{4.54}
\end{eqnarray}
Applying the Foldy-Wouthuysen transformation (\ref{3.FWT}) we
obtain for the matrix elements $\langle{\bi x}|\rme^{-t\gamma_0 {H}_{\rm D}({\hat {\bi p}})}| {\bi
x}'\rangle $ the path integral
\begin{eqnarray}
 \langle{\bi x}|\rme^{-t\gamma_0 {H}_{\rm D}({\hat {\bi p}})}| {\bi
x}'\rangle
&=& \int_{\mathbb{R}^{D+1}}
\frac{\rmd^{D+1} p}{(2\pi)^{D+1}} \ \! \rme^{-\rmi p(x - x')} \
\! V_{{\bi p}}\left(\rmi c p_0 + H_{\bi p} \gamma_0
\right)V^{-1}_{{\bi p}} \
\!\frac{1}{p^2+m^2c^2}\nonumber \\[1mm]
\comment{
&=& \int_{\mathbb{R}^{D+1}} \frac{\rmd^{D+1} p}{(2\pi)^{D+1}} \ \!
\rme^{-\rmi p(x - x')} \ \! (\rmi c p_0 + c\gamma^0\bm{\gamma}\cdot
{\bi p}   +
\gamma^0 m c^2) \ \!\frac{1}{p^2+m^2c^2}\nonumber \\[1mm]
}
&=& \frac{\gamma_{_{\rm{E}}}^0}{c}\int_{\mathbb{R}^{D+1}} \frac{\rmd^{D+1} p}{(2\pi)^{D+1}}\ \!
\rme^{-\rmi p(x - x')} \ \! \frac{\rmi/\!\!\!p_{_{\rm{E}}} + m c}{p^2 + m^2c^2}
\,\, . \label{IV51a}
\end{eqnarray}
To obtain the last line we have used the simple matrix identity
$V_{{\bi p}}\left(\rmi c p_0 + H_{\bi p} \gamma_0
\right)V^{-1}_{{\bi p}}=
\rmi c p_0 + c\gamma^0\bm{\gamma}\cdot
{\bi p}   +
\gamma^0 m c^2$, and introduced the euclidean gamma matrices
$\gamma^0_{_{\rm{E}}} \equiv \gamma^0$ and $\bm{\gamma}_{_{\rm{E}}}
\equiv \rmi\bm{\gamma}$ fulfilling the Clifford algebra
$\{\gamma^{\mu}_{_{\rm{E}}}, \gamma^{\nu}_{_{\rm{E}}}\} = 2
\delta^{\mu \nu}$. Similarly $/\!\!\!p_{_{\rm{E}}} \equiv
\gamma^0_{_{\rm{E}}} p^0 + \bm{\gamma}_{_{\rm{E}}}\!\cdot {\bi p}$.

The euclidean amplitude (\ref{IV51a}) satisfies the Fokker-Planck-like equation
analogous to (\ref{2.2aab}):
\begin{eqnarray}
(\partial_{t} + {H}_{\rm D})
\langle{\bi x}|\rme^{-t\gamma_0 {H}_{\rm D}({\hat {\bi p}})}| {\bi
x}'\rangle
 \ =
\ \delta(t-t') \delta^{(D)}(\bi x-\bi x')\, .
\label{3.3aa}\end{eqnarray}
By multiplying this with $\gamma^0_{_{\rm{E}}}$, and
defining $ \tilde{P}( {\bi x},t|{\bi x}',t')\equiv
\langle{\bi x}|\rme^{-t\gamma_0 {H}_{\rm D}({\hat {\bi p}})}| {\bi x}'\rangle
c \gamma^0_{_{\rm{E}}}$, we obtain
the covariant expression
\begin{eqnarray}
(\gamma^0_{_{\rm{E}}} \partial_{ct} +{c^{-1}}\gamma^0_{_{\rm{E}}} {H}_{\rm D}) \tilde{P}( {\bi x},t|{\bi
x}',t')  \ = \ (\rmi/\!\!\!p_{_{\rm{E}}} + mc)
 \tilde{P}(
{\bi x},t|{\bi x}',t') \ = \  \delta ^{(D+1)}(x-x')\, ,
\label{3.3ab}\end{eqnarray}
showing that $\tilde{P}( {\bi x},t|{\bi x}',t')$ is
the  Green function for the euclidean Dirac equation.

Let us finally emphasize the well-known fact that using only the positive
energies in relativistic path integrals such as
Eq.~(\ref{1.4a}) leads immediately to pathologies, such as
non-conservation of probability, lack of Zitterbewegung~\cite{zitt},
loss of relativistic invariance~\cite{hartle:01} and, of course,
the inability to find the correct Klein-Gordon
propagator. These difficulties do not arise when the {\em full
matrix structure} of the Weibull distribution is taken into account.
Such a matrix structure takes complete care of both particles and
antiparticles and it highlights the key role of the
Feynman-Stuckelberg boundary condition.


\section{Reparametrization freedom \label{SEc4}}

Let us now turn to the case of a general reparametrization
invariance. In this connection it is instructive to observe various
degrees of freedom in the representation (\ref{1.4}) of the
conditional probability (\ref{1.4a}). First we note that by
substituting $v \mapsto v/a^2$ we obtain the identity
\begin{eqnarray}
&&\mbox{\hspace{-9mm}}\int_{{\bi x}(0)= {\bi x}'}^{{\bi x}(t) = {\bi
x}''} {\mathcal{D}}{\bi x} \frac{\mathcal{D} {\bi p}}{(2\pi)^D} \ \!
\exp\left\{\int_{0}^{t} \rmd \tau \ \!\left[\rmi {\bi p}\cdot
\dot{\bi x} \ - \
c\sqrt{{\bi p}^2 + m^2 c^2}\right]\right\} \nonumber \\[2mm]
&&\mbox{\hspace{-4mm}}=   \int_{0}^{\infty}\!\rmd v \ \!
\omega(v,a,t) \int_{{\bi x}(0)= {\bi x}'}^{{\bi x}(t) = {\bi x}''}
{\mathcal{D}}{\bi x}  \  \! \frac{\mathcal{D} {\bi p}}{(2\pi)^D} \ \!
\exp\left\{\int_{0}^{t} \rmd \tau\ \!\left[\rmi {\bi p}\cdot
\dot{\bi x}\ - \ \frac{v}{a^2} ({\bi p}^2 c^2 + m^2
c^4)\right]\right\}, \label{SEc4.1a}
\end{eqnarray}
where $\omega(v,a,t)$ is the Weibull distribution of order $a$
in
Eq.~(\ref{weib.aa}). The right-hand side can be further
integrated
functionally over ${\bi p}$ to become
\begin{eqnarray}
&&\mbox{\hspace{-9mm}} \int_{0}^{\infty}\!\rmd v \ \! \omega(v,a,t)
\int_{{\bi x}(0)= {\bi x}'}^{{\bi x}(t) = {\bi x}''}  {\mathcal{D}}{\bi x}
\  \! \exp\left\{-\int_{0}^{t}
\rmd \tau \left[\frac{a^2}{4 c^2 v}\ \!\dot{\bi x}^2  \ + \ \frac{c^2 v}{a^2}
\ \! m^2c^2 \right] \right\} \nonumber \\[2mm]
&&\mbox{\hspace{-4mm}}=\int_{0}^{\infty} \rmd L \ \ \!\frac{ct \
\!\rme^{-c^2t^2/2L}}{\sqrt{2\pi L^3}} \! \int_{{\bi x}(0)= {\bi
x}'}^{{\bi x}(t) = {\bi x}''}  {\mathcal{D}}{\bi x} \  \!
\exp\left\{-\int_{0}^{\lambda} \rmd \bar{\lambda}  \left[ \frac{1}{2
e} \ \![{\bi x}'(\bar{\lambda})]^2  \ + \ \frac{e}{2} \ \! m^2c^2
\right] \right\}  . \label{SEc4.1b}
\end{eqnarray}
%
%
Here we have defined a new variable $\bar{\lambda}\equiv\tau\,
2c^2v/a^2e$, so that the length of a particle orbit
is $L \equiv \int_0^{\lambda} \rmd \bar{\lambda} \ \! e$. In this
expression, $e$ may be viewed as a constant ``einbein", i.e., a
square root of the intrinsic metric along the worldline. As in
Section~\ref{SEc2b} we can rewrite (\ref{SEc4.1b}) in a relativistic
form by utilizing an auxiliary Gaussian path integral for $x_0$
similar to (\ref{3.38a}), as
\begin{eqnarray}
 \partial_t \int_{x_0(0) = 0}^{x_0(\lambda) = ct}  {\mathcal{D}}x_0 \ \!
 \exp\left\{-\int_{0}^{\lambda}
\rmd \bar{\lambda}  \ \! \frac{1}{2 e} \ \![{x_0'(\bar{\lambda})}]^2 \right\}
\ = \  - c^2 t \sqrt{\frac{1}{2\pi L^3}} \ \! \rme^{-c^2t^2/2L}\, .
\end{eqnarray}
With this we can rewrite the right-hand side of (\ref{SEc4.1a}) as
\begin{eqnarray}
-\ \! \frac{\partial_t}{c}  \int_{0}^{\infty}  \!\rmd  L \ \!
\int_{x(0)= { x}'}^{{ x}(\lambda) = {x}''}  {\mathcal{D}}{ x} \  \!
\exp\left\{-\int_{0}^{\lambda} \rmd \bar{\lambda}  \left[ \frac{1}{2
e} \ \![{x_{\mu}'(\bar{\lambda})}]^2 \ + \ \frac{e}{2} \ \! m^2c^2
\right] \right\}.
\end{eqnarray}
Analogous steps to those in Section~\ref{SEc2b}  allow us to to
find for the Klein-Gordon propagator (\ref{2.6ax})  the worldline
representation
\begin{eqnarray}
&&\mbox{\hspace{-9mm}}\int_{0}^{\infty}  \!\rmd  L \ \! \int_{x(0)=
{ x}'}^{{ x}(\lambda) = {x}''}  {\mathcal{D}}{ x} \  \! \exp\left\{-\int_{0}^{\lambda}
\rmd \bar{\lambda}  \left[ \frac{1}{2 e} \ \![{x_{\mu}'(\bar{\lambda})}]^2
\ + \ \frac{e}{2} \ \! m^2c^2 \right] \right\}\nonumber \\[2mm]
&&\mbox{\hspace{4mm}} = \
\int_{\mathbb{R}^{D+1}} \frac{\rmd^{D+1} p}{(2\pi)^{D+1}} \ \!
\frac{\rme^{-\rmi p(x'' - x')}}{p^2 \ + \ m^2c^2}\,
  .\label{SEc4.4}
\end{eqnarray}

So the different choices of the parameter $a$ of the Weilbull smearing
distribution correspond to different constant einbeins $e$ in the
worldline representations of the Klein-Gordon propagator.
\comment{
Note that
because $a$ is a constant $e$ is not a dynamical variable and hence
there is no gauge fixing required (there is no integration over
$e$). Notice also that all representations (\ref{SEc4.4}) are
isomorphic to Polyakov gauge as they can be obtain from it through a
{\em constant} rescalling of a label time.
}

The freedom of choice of $e$ in (\ref{SEc4.4}) can be generalized
further to a full gauge  freedom, i.e., a freedom to change the
worldline parametrization. However, this cannot be
done straightforwardly just by assuming that $a$
depends on $\bar \lambda$. This is because the Hamiltonian we wish to smear out
would become explicitly ``time-dependent'' [see Eq.~(\ref{SEc4.1a})],
and for such cases our superstatistics argument~\cite{JK:08} is not valid.
It is, however, not difficult to tackle the problem indirectly. To
see this we use a simple identity for a functional $\delta$-function~\cite{Z-J:04}:
let $e(\bar{\lambda})$ be a dynamical variable and let
$F_{\bar{\lambda}}(e) = 0$ be a system of equations that for each
 $\bar{\lambda}$ provides a constant solution $e_{\rm s}$.
Let, in addition, $F_{\bar{\lambda}} = F_{\bar{\lambda}}(e)$ be
a one-to-one map in some neighborhood of $F_{\bar{\lambda}} = 0$ which
can be inverted to $e(\bar{\lambda}) \equiv e_{_{\!\bar{\lambda}}} =
e_{_{\!\bar{\lambda}}}(F)$. Any functional
$G[e_{\rm s}]$ can be then written as
\begin{eqnarray}
G[e_{\rm s}] \ &=& \ \int \left[ \prod_{\bar{\lambda}}\rmd F_{_{\!\bar{\lambda}}}
\ \! \delta(F_{_{\!\bar{\lambda}}})\right]G[e(F)]\ = \ \int \left[\prod_{\bar{\lambda}}
\rmd e_{_{\bar{\lambda}}} \ \!\delta(F_{_{\!\bar{\lambda}}}(e))\right] \mathcal{J}(e) G[e] \nonumber \\[2mm]
&\equiv& \ \int {\mathcal{D}} e \ \! \delta[F(e)]\ \! \mathcal{J}(e)
G[e]\, ,
\end{eqnarray}
with the functional Jacobian
\begin{eqnarray}
\mathcal{J}(e) \ = \ \det F_{{\bar{\lambda}} {{\bar{\lambda}}}'},
\;\;\;\;\;\;\;\;\;\; F_{\bar{\lambda} {\bar{\lambda}}'} \ = \
\frac{\partial F_{_{\!\bar{\lambda}}}}{\partial e(\bar{\lambda}')}\, .
\end{eqnarray}
By setting
\begin{eqnarray}
G[e_{\rm s}] \ = \  \exp\left\{-\int_{0}^{\lambda}
\rmd \bar{\lambda}  \left[ \frac{1}{2 e_{\rm s}} \ \![{x_{\mu}'(\bar{\lambda})}]^2
\ + \ \frac{e_{\rm s}}{2} \ \! mc^2 \right] \right\},
\end{eqnarray}
we can rewrite (\ref{SEc4.4}) in the form
\begin{eqnarray}
&&\mbox{\hspace{-2mm}}\int_{0}^{\infty} \! \!\rmd  L \ \! \int \!\!{\mathcal{D}} e \ \!
\delta[F(e)]\ \! \mathcal{J}(e) \ \! \int_{x(0)= { x}'}^{{ x}(\lambda) = {x}''} \!\!
{\mathcal{D}}{ x} \  \! \exp\left\{-\int_{0}^{\lambda}
\rmd \bar{\lambda}  \left[ \frac{1}{2 e(\bar{\lambda})} \ \![{x_{\mu}'(\bar{\lambda})}]^2
\ + \ \frac{e(\bar{\lambda})}{2} \ \! m^2c^2 \right] \right\}\nonumber \\[2mm]
&&\mbox{\hspace{-2mm}}= \ \int_{\mathbb{R}^{D+1}} \frac{\rmd^{D+1} p}{(2\pi)^{D+1}} \ \!
\frac{\rme^{-\rmi p(x'' - x')}}{p^2 \ + \ m^2c^2}\, . \label{65}
\end{eqnarray}
The action in (\ref{65}) is clearly {\em reparametrization-invariant} under
 $\bar{\lambda} \mapsto \bar{\lambda}' = f(\bar{\lambda})$,
if we transform
\begin{eqnarray}
x^{\mu}(\bar{\lambda})\ \mapsto \ \tilde{x}^{\mu}(\bar{\lambda}) \ = \ x^{\mu}(f^{-1}(\bar{\lambda})),
\;\;\;\;\; e(\bar{\lambda}) \ \mapsto \ e'(\bar{\lambda}) \ = \
\frac{\rmd f^{-1}(\bar{\lambda})}{\rmd \bar{\lambda}}e(f^{-1}(\bar{\lambda}))\, .\label{67aa}
\end{eqnarray}
Here $f(\bar{\lambda})$  is an arbitrary monotonically increasing function of $\bar{\lambda}$.
A general gauge fixing, say $\tilde{F}(e) = 0$, can be now implemented by performing
the change of the einbein variable $e\mapsto e'$ via $e = F^{-1}\mbox{\footnotesize{$\circ$}}\ \!\tilde{F}(e')$.
As a consequence of the rules of functional differentiation we have
\begin{eqnarray}
\mathcal{D} e \ \! \delta[F(e)]\ \! \mathcal{J}(e) \ &=& \ \mathcal{D} e' \ \! \det\!
\left[\frac{\partial (F^{-1}\mbox{\footnotesize{$\circ$}}\ \!\tilde{F})}{\partial e'}\right]\delta[\tilde{F}(e')] \ \!
\det\!\left[\frac{\partial(F \mbox{\footnotesize{$\circ$}}F^{-1}\mbox{\footnotesize{$\circ$}}\ \!
\tilde{F})}{\partial(F^{-1}\mbox{\footnotesize{$\circ$}}\ \!\tilde{F})}\right] \nonumber \\[2mm]
&=& \ \mathcal{D} e' \ \! \delta[\tilde{F}(e')]\ \! \tilde{\mathcal{{J}}}(e')\, ,
\end{eqnarray}
where $\delta[\tilde{F}(e')] = \prod_{_{\bar{\lambda}'}} \delta(\tilde{F}_{_{\bar{\lambda}'}}(e'))$,
and the functional Jacobian $\tilde{\mathcal{J}}$ has the form
\begin{eqnarray}
\tilde{\mathcal{J}}(e') \ = \ \det \tilde{F}_{{\bar{\lambda}} {{\bar{\lambda}}}'},
\;\;\;\;\;\;\;\;\;\; \tilde{F}_{\bar{\lambda} {\bar{\lambda}}'} \ = \
\frac{\partial \tilde{F}_{_{\!\bar{\lambda}}}}{\partial e'(\bar{\lambda}')}\, .
\end{eqnarray}
Note also that due to einbein identity: $\rmd \bar{\lambda}\,
e(\bar{\lambda}) = \rmd \bar{\lambda}'\, e'(\bar{\lambda}')$ (see, e.g., Ref.~\cite{PI,Polchinski})
the action in (\ref{65}) changes to
\begin{eqnarray}\!\!
\int_{0}^{\lambda}
\rmd \bar{\lambda}  \left[ \frac{1}{2 e'(\bar{\lambda})} \ \![{\tilde{x}_{\mu}'(\bar{\lambda})}]^2
\ + \ \frac{e'(\bar{\lambda})}{2} \ \! mc^2 \right] .
\end{eqnarray}

We can now relabel $e'$ back to $e$ and $\tilde{x}^{\mu}$ back to ${x}^{\mu}$, and
write the left-hand side of (\ref{65}) in the gauge-fixed form
%
\begin{eqnarray}
\mbox{\hspace{-2mm}}\int_{0}^{\infty} \! \!\rmd  L  \int
\!\!{\mathcal{D}} e \ \! \delta[\tilde{F}(e)]\ \!
\tilde{\mathcal{J}}(e)  \int_{x(0)= { x}'}^{{ x}(\lambda) =
{x}''} \!\! {\mathcal{D}}{ x} \  \! \exp\left\{-\int_{0}^{\lambda}
\rmd \bar{\lambda}  \left[ \frac{1}{2 e(\bar{\lambda})} \
\![{x_{\mu}'(\bar{\lambda})}]^2 \ + \ \frac{e(\bar{\lambda})}{2} \
\! m^2c^2 \right] \right\} ,\label{67}
\end{eqnarray}
with a particle orbit length
\begin{eqnarray}
L \ = \ \int_{0}^{\lambda} \rmd \bar{\lambda} \ \! e(\bar{\lambda})\, .
\end{eqnarray}
The reader may notice that the gauge $e\equiv {\rm const.}$ is
recovered by setting $\tilde{F}_{\bar{\lambda}}(e) =
e(\bar{\lambda}) -e$.

Let us finally observe that (\ref{67}) can be rewritten as
\begin{eqnarray}
&&\int_{0}^{\infty} \! \!\rmd  L \ \! \int \!\!{\mathcal{D}} e \ \!
\delta[\tilde{F}(e)]\ \! \tilde{\mathcal{J}}(e) \ \!
\frac{\exp[-(x_{0}'' - x_0')/2L]}{\sqrt{2\pi} L}
\nonumber \\[2mm]
&&\mbox{\hspace{15mm}}\times \ \int_{x(0)= { x}'}^{{ x}(\lambda) =
{x}''} \!\! {\mathcal{D}}{\bi x}\  \! \frac{\mathcal{D} {\bi
p}}{(2\pi)^D} \ \! \exp\left\{\int_0^{\lambda}  \d \bar{\lambda}
\left[\rmi {\bi p}\cdot {\bi x} - \frac{e(\bar{\lambda})}{2}({\bi
p}^2 + m^2c^2) \right] \right\}\! , \label{68}
\end{eqnarray}
which indicates that the smearing-distribution functional
corresponding to the einbein $e(\bar{\lambda})$ reads
\begin{eqnarray}
\varrho[e; x_0'', x_0'] \ = \
\delta[\tilde{F}(e)]\tilde{\mathcal{J}}(e) \ \! \frac{\exp[-(x_{0}''
- x_0')/2L]}{\sqrt{2\pi} L}\, . \label{69}
\end{eqnarray}
In deriving (\ref{68}) we have used the fact that
\begin{eqnarray}
\int_{x_0(0)= \ \!x_0'}^{x_0(\lambda) = \ \!x_0''} {\mathcal{D}}
\left(\frac{x_0}{\sqrt{e}}\right) \! \exp\left\{-\int_{0}^{\lambda} \rmd
\bar{\lambda}\ \! \frac{1}{2e(\bar{\lambda})} \
\![x_0'(\bar{\lambda})]^2 \right\}
 \ = \  \frac{\exp\!\left[-{(x_0'' - x_0')^2}/{2 L} \right]}{\sqrt{2\pi L}}\, ,  \label{70}
\end{eqnarray}
with
(see, e.g., Ref.~\cite{PI,polyakov:87})
\begin{eqnarray}
{\mathcal{D}}\!\left(\frac{{x}_0}{\sqrt{e}}\right)\ \equiv \
\sqrt{\frac{1}{2\pi \varDelta \bar{\lambda}_0 e_{_{\bar{\lambda}_0}}}} \ \! \prod_{_{\bar{\lambda}_i}} \frac{\rmd{{
x}_0}({\bar{\lambda}_i})}{\sqrt{2\pi \varDelta \bar{\lambda}_i {{e}}_{_{\bar{\lambda}_i}}}},
\end{eqnarray}
%
and the identity~\cite{Grosche:98}:
\begin{eqnarray}
&&\mbox{\hspace{-15mm}}\int_{{\bi x}(0)= {\bi x}'}^{{\bi x}(\lambda)
= {\bi x}''} {\mathcal{D}}{\bi x}  \  \! \frac{\mathcal{D} {\bi
p}}{(2\pi)^D} \ \! \exp\left\{\int_{0}^{\lambda} \rmd \bar{\lambda}\
\!\left[\rmi {\bi p}\cdot \dot{\bi x}\ - \
\frac{e(\bar{\lambda})}{2} {\bi
p}^2\right]\right\}\nonumber \\[2mm]
&&\mbox{\hspace{15mm}}= \ \int_{{\bi x}(0)= {\bi x}'}^{{\bi
x}(\lambda) = {\bi x}''} {\mathcal{D}}\!\left(\frac{{\bi
x}}{{e}^{D/2}}\right) \! \exp\left\{-\int_{0}^{\lambda} \rmd
\bar{\lambda}\ \! \frac{1}{2e(\bar{\lambda})} \ \![{\bi
x}'(\bar{\lambda})]^2 \right\}\, .
 \label{71}
\end{eqnarray}
In the above definition the interval $0\leq \bar{\lambda}\leq
\lambda$ is split into not necessarily equal slices $\varDelta
\bar{\lambda}_i$ in order to preserve  the integration measure under
the reparametrization transformations (\ref{67aa}).  In particular, while $\varDelta
\bar{\lambda}_i$ are non-equal slices, $\varDelta
\bar{\lambda}_i\ \!{{e}}_{_{\bar{\lambda}_i}}$ is constant for all $i$
because ${{e}}^2(\bar{\lambda})$ is the one-dimensional
version of the  metric ``tensor'' along the path.

\section{Connection with emergent relativity \label{SEc5a}}

The identity (\ref{SEc4.1a}) can be interpreted in yet another
interesting way. To this end, we rewrite Eq.~(\ref{SEc4.1a}) as
\begin{eqnarray}
&&\mbox{\hspace{-10mm}}\int_{{\bi x}(0)= {\bi x}'}^{{\bi x}(t) =
{\bi x}''} \!\!{\mathcal{D}}{\bi x} \frac{\mathcal{D} {\bi
p}}{(2\pi)^D} \ \! \exp\left\{\int_{0}^{t} \!\!\rmd \tau \
\!\left[\rmi {\bi p}\cdot \dot{\bi x} \ - \
c\sqrt{{\bi p}^2 + m^2 c^2}\right]\right\} \nonumber \\[2mm]
&&\mbox{\hspace{-10mm}}= \  \int_{0}^{\infty}\!\!\rmd \tilde{m} \ \!
\sqrt{\frac{c^2t}{2\pi \tilde{m}}}\ \! \rme^{-tc^2(\tilde{m} -
m)^2/2\tilde{m}}  \int_{{\bi x}(0)= {\bi x}'}^{{\bi x}(t) = {\bi
x}''} \!\!{\mathcal{D}}{\bi x}  \ \! \frac{\mathcal{D} {\bi
p}}{(2\pi)^D} \ \!  \exp\left\{\int_{0}^{t} \!\!\rmd \tau\
\!\left[\rmi {\bi p}\cdot \dot{\bi x} -  \frac{{\bi p}^2}{2
\tilde{m}}  -  m c^2
\right]\right\}\nonumber \\[2mm]
&&\mbox{\hspace{-10mm}}= \  \int_{0}^{\infty}\!\!\rmd \tilde{m} \ \!
f_{\frac{1}{2}}\!\left(\tilde{m}, tc^2, tc^2m^2\right) \int_{{\bi
x}(0)= {\bi x}'}^{{\bi x}(t) = {\bi x}''} \!\!{\mathcal{D}}{\bi x} \
\! \frac{\mathcal{D} {\bi p}}{(2\pi)^D} \ \! \exp\left\{\int_{0}^{t}
\!\!\rmd \tau\ \!\left[\rmi {\bi p}\cdot \dot{\bi x} -  \frac{{\bi
p}^2}{2 \tilde{m}}  -  m c^2 \right]\right\}, \label{SEc5.1a}
\end{eqnarray}
where
\begin{eqnarray}
f_p(z,a,b) \ = \ \frac{(a/b)^{p/2}}{2K_p(\sqrt{a b})} \ \! z^{p-1} \
\!\rme^{-(az + b/z)/2}\, ,
\end{eqnarray}
(with $K_p=$ modified Bessel function of the second kind)
is the {\em generalized inverse Gaussian}
distribution~\cite{jorgensen} (known also as {\em Sichel}'s distribution).
From the form of the Hamiltonian in
(\ref{SEc5.1a}) we see that the mass $\tilde{m} $ plays role of the
ordinary Newtonian mass which takes on continuous values distributed
according to distribution $f_{\frac{1}{2}}\!\left(\tilde{m}, tc^2,
tc^2m^2\right)$ with the expectation value  $\langle \tilde{m}
\rangle = m + 1/tc^2$. Relation (\ref{SEc5.1a}) can then be given
the following heuristic interpretation: Single-particle relativistic
theory might be viewed as a single-particle {\em non\/}-relativistic theory
whose Newtonian mass $\tilde{m}$ (which is not invariant under
Lorentz transformations) is a fluctuating parameter
whose average approaches the true relativistic
mass $m$ in the large $t$ limit. In view of results of Ref.~\cite{JK:08} we can more formally
state that a stochastic process described by the Kramers-Moyal
equation with the relativistic Hamiltonian $c\sqrt{{\bi p}^2 + m^2
c^2}$ is equivalent to a doubly stochastic process in which the
fast-time dynamics of a free non-relativistic particle (Brownian
motion) is coupled with the long-time dynamics describing
fluctuations of particle's Newtonian mass. On more speculative vein,
one can fit the above observation into currently much debated
``emergent (special) relativity''. The emergent relativity tries to
view a special theory of relativity as a theory that statistically
emerges from a deeper (essentially non-relativistic) level of
dynamics. It dates back to works of D.~Bohm~\cite{bohm:96,bohm:93}
in early 50's, but it received a real boost  with the advancement of
quantum-gravity approaches. In recent years it has appeared under
various disguises in quantum-gravity models based on spacetime foam
pictures~\cite{garay:98},  in loop quantum gravity
models~\cite{gambini:99,alfaro:00},  in non-commutative geometry
models~\cite{camelia:00,susskind:00,camelia:01,douglas:01} or in
black-hole physics~\cite{jacobson:08}.

At a strictly phenomenological level, one can understand
fluctuations of the Newtonian mass as originating from the idea that
the medium in which propagation occurs (``spacetime'') involves some
sort of ``granularity'' (usually considered in quantum gravity
models). On the basis of experience with condensed-matter systems,
one can expect that granularity of the medium might lead to
corrections in the local dispersion relation and hence to
modifications in local effective mass of a particle.

Suppose, now, that on the fast-time level a non-relativistic
particle propagates through grains with a different local
$\tilde{m}$ in each grain (e.g., crystalline grains with a different
local lattice structures or lattice spacings). Assume that the
probability of the distribution of $\tilde{m}$ in various grains is
$f_{\frac{1}{2}}\!\left(\tilde{m}, tc^2, tc^2m^2\right)$. Because
the fast-time scale motion is brownian, the local probability
density matrix (PDM) conditioned on some fixed $\tilde{m}$ in a
given grain is Gaussian:
\begin{eqnarray}
\hat{\rho}({\bi p},t|\tilde{m}) \ \propto \  \rme^{- t\hat{\bi
p}^2/2\tilde{m}}\, .
\end{eqnarray}
The joint PDM is then $\hat{\rho}({\bi p},t;\tilde{m}) =
f_{\frac{1}{2}}\!\left(\tilde{m}, tc^2,
tc^2m^2\right)\hat{\rho}({\bi p},t|\tilde{m})$ and the marginal PDM
describing the mass-averaged (i.e. long-time) behavior of the
particle is
\begin{eqnarray}
\hat{\rho}({\bi p},t) = \int_{0}^{\infty} \rmd \tilde{m} \ \! f_{\frac{1}{2}}\!\left(\tilde{m}, tc^2,
tc^2m^2\right)\hat{\rho}({\bi p},t|\tilde{m})\, .
\end{eqnarray}
Local matrix element in the position basis, i.e. $\langle {\bi x}|
\hat{\rho}({\bi p},t-t')|{\bi x}' \rangle$, then corresponds to
transition probability $P({\bi x}, t| {\bi x}', t')$ which has the
Newton-Wigner path-integral representation (\ref{1.4a}).

It should be noted that these conclusions extend also to less
trivial situations. One may, for instance, consider the Klein-Gordon
or Dirac particle coupled  to an external electromagnetic field
$A_{\mu}({\bi x},t)$ and to a scalar potential $V({\bi x},t)$. In
such a case Dirac's Hamiltonian is
\begin{eqnarray}
{H}_{\rm D}^{A,V} \ = \ c\gamma_0 \bm{\gamma} \cdot ({\bi p} -  e {\bi
A}/c) \ + \ \gamma_0 (m c^2 + V) \ + \ e A_0\, . \label{VIba}
\end{eqnarray}
and the Feshbach-Villars Hamiltonian reads
\begin{eqnarray}
{H}_{\rm FV}^{A,V} \ = \ (\sigma_3 + \rmi \sigma_2)\frac{1}{2m}({\bi p} - e
{\bi A}/c)^2 \ + \ \sigma_3 (m c^2 + V) + e A_0\, , \label{VIaa}
\end{eqnarray}
For the purpose of illustrating our point we will focus on an
electron in a magnetostatic field ${\bi B} = \mbox{rot} {\bi A }$
with  $V =0$. In this case, there exists a $4\times 4$
Foldy-Wouthuysen-like transformation~\cite{case:54,eriksen:58} that
brings Dirac's Hamiltonian (\ref{VIba}) to a quasi-diagonal form:
\begin{eqnarray}
{H}_{\rm D}^{A,V}({\bi p}, {\bi x}) \ = \ V_{{\bi p}, {\bi x}}
{H}_{\rm diag}^{A,V}({\bi p}, {\bi x})
V_{{\bi p}, {\bi x}}^{-1}\, , \label{6.82a}
\end{eqnarray}
where
\begin{eqnarray}
{H}_{\rm diag}^{A,V}({\bi p}, {\bi x}) \ = \
\gamma_{0}\sqrt{c^2({\bi p} -  e {\bi A}/c)^2 + m^2c^4 - e\hbar c
{\bi B}\cdot {\bi \varSigma}}, \;\;\;\;\;\;\; \varSigma_i \ = \
\frac{i}{4} \epsilon_{ijk}[\gamma_j, \gamma_k]\, , \label{6.83a}
\end{eqnarray}
and
\begin{eqnarray}
V_{{\bi p}, {\bi x}} \ = \ \rme^{-\rmi S_{{\bi p}, {\bi x}}},
\;\;\;\;\;\; S_{{\bi p}, {\bi x}} \ = \ - \mbox{$\frac{1}{2}$}
\arctan\left( \frac{\rmi\bm{\gamma} \cdot (c {\bi p} - e {\bi
A})}{mc^2} \right).
\end{eqnarray}
For ${{\bi A}} = 0$ this reduces to the Foldy-Wouthuysen
transformation (\ref{3.1aa}) as one can easily check by comparing
respective Taylor series. Our analysis will further simplify when
the magnetic field is also spatially constant. In this case the
vector potential can be taken as: $A_x = -B y$ and $A_y = A_z = 0$
(the $z$-axis is chosen to be in the ${\bi B}$ direction, $B_z
\equiv B$) and then
\begin{eqnarray}
{H}_{\rm diag}^{A,V}({\bi p}, {\bi x}) \ &=& \
\gamma_{0}\sqrt{c^2({p}_x + e B y/c)^2 + c^2 p_y^2 + c^2 p_z^2 +
m^2c^4 - e\hbar c B\varSigma_z}\nonumber \\[2mm]
&=& \  \sigma_3\otimes \sqrt{c^2 {\bi p}^2 + e^2B^2 y^2 - e c
B(\hbar \sigma_3 - 2y p_x)+ m^2c^4 }\, .
\end{eqnarray}
Let us observe that the
latter is already in a diagonal form. Following the procedure from
Sections~\ref{SEc2b} and \ref{SEc3}, the key object to be evaluated
is the imaginary-time propagator
\begin{eqnarray}
&&\hspace{-10mm}\langle {\bi x} |\rme^{-(t-t'){H}_{\rm
diag}^{A,V}({\bi p}, {\bi
x}) } |{\bi x}' \rangle  \nonumber \\[2mm]
&& \hspace{-8mm} =  \ \int_{0}^{\infty} \rmd v \
\!\omega(v,t)\otimes \int_{{\bi x}(t') = {\bi x}'}^{{\bi x}(t) =
{\bi x}} {\mathcal{D}}{\bi x} \ \!\frac{{\mathcal{D}}{\bi
p}}{(2\pi)^D} \,
 \rme^{ \int_{t'}^{t}\rmd \tau \left\{ \rmi {\bi p}\cdot\dot{\bi x}\ -
\ v [{\bi p}^2 c^2  + e^2B^2 y^2 - e c B(\hbar \sigma_3 - 2y p_x) +
m^2c^4]\right\} }\nonumber \\[2mm]
&& \hspace{-8mm} =  \ \int_{0}^{\infty} \rmd \tilde{m} \ \!
f_{\frac{1}{2}}(\tilde{m}, tc^2, tc^2m^2) \otimes \int_{{\bi x}(t')
= {\bi x}'}^{{\bi x}(t) = {\bi x}} {\mathcal{D}}{\bi x} \
\!\frac{\mathcal{D}{\bi p}}{(2\pi)^D} \  \! \exp\left\{
\int_{t'}^{t}\!\!\rmd \tau \left[\rmi {\bi p}\cdot\dot{\bi x} -
H_{\rm SP}  - mc^2 \right] \!\right\} \!,
\end{eqnarray}
where $H_{\rm SP}$ corresponds to the nonrelativistic Hamiltonian for a
particle in a constant uniform magnetic field (Schr\"{o}dinger-Pauli
Hamiltonian)
\begin{eqnarray}
H_{\rm SP} \ = \ \frac{1}{2 \tilde{m}}\left[ \left(p_x + \frac{e}{c}
B y \right)^{\! 2} \  + \ p_y^2 \ + \ p_z^2 \right] \ - \ \mu_{\rm
B} B \sigma_3 \, ,
\end{eqnarray}
with $\mu_{\rm B} = e\hbar/2\tilde{m}c$ representing the Bohr magneton.

Note, in particular, that the smearing distributions $\omega$ and
$f_{\frac{1}{2}}$ stay the same as in the free-particle case (cf.
Eq.~(\ref{2.33a}) and Eq.~(\ref{SEc5.1a})). Diagonalization
analogous to (\ref{6.82a})-(\ref{6.83a}) can be performed also for
charged spin-$0$ particles, such as, e.g, $\pi^{\pm}$
mesons~\cite{case:54}.

Two points hinder this program to be carried further in a full
generality: firstly, general $x$ dependence of $A_{\mu}$ and $V$
leads to a notorious ordering problem. Secondly, and most
importantly, transformation that would bring the Hamiltonian into a
form where the positive and negative parts are explicitly separated
is no longer possible for a general interaction. This last point
makes impossible to carry over straightforwardly our reasonings from
Sections~\ref{SEc2b} and \ref{SEc3}.

\section{Concluding remarks \label{SEc5}}

In this paper we have extended an earlier paper~\cite{JK:08}
on superstatistics to a calculation
of the worldline representations of Feynman propagators for
spin-$0$ an spin-$\frac{1}{2}$ particles via a superstatistical average
of non-relativistic single particle paths.
For conceptual reasons we have found it useful to describe the spin-$0$ particle by the
less-known Feshbach-Villars rather than the usual Klein-Gordon equation.
The Feshbach-Villars representation casts the Klein-Gordon equation
into two equations, both of which are first order in time.
Because of this first-order nature we could use
the Feynman-Kac formula to set
up the path-integral representation for the corresponding Feynman's
propagator.  Two-component nature of the wave
function, in addition, allowed to treat the positive and negative energy
solutions on equal footing and easily accommodated the Feynman-Stuckelberg
boundary condition. This considerably facilitated our calculations.
Although we have discussed only spin-$0$ and spin-$\frac{1}{2}$
particles, the method could have been also employed to discuss the
Proca equation for spin-$1$ particle. This is because for Proca's
Hamiltonian one can find an analogous diagonalisation  as in
spin-$0$ and spin-$\frac{1}{2}$ cases~\cite{case:54}.

From the superstatistics viewpoint, the relations (\ref{1.4})
and (\ref{SEc4.1a}), as well as their matrix generalizations
(\ref{eq.26}) and (\ref{53}), belong to the inverse
$\chi^2$-superstatistics universality class~\cite{beck:05,beck:09}.
This is a rather interesting result, in particular
when we realize that Weibull's smearing distribution was unambiguously forced upon us by
requiring that the smeared Gaussian Markovian process (i.e.,
non-equilibrium Markovian processes) should be identical with the
Newton-Wigner Markovian process~\cite{JK:08}, i.e., a Markovian
process with the square-root Hamiltonian $H_{\bi p}$. In fact, in
Ref.~\cite{JK:08} we have shown that mere requirement that a smeared
Markovian process should be again Markovian process naturally
resulted in both inverse $\chi^2$- and $\chi^2$-superstatistics
universality class. In this view it is plausible to conjecture that
superstatistics equivalence classes are closely related to smearing
distributions in non-equilibrium Markovian systems.

In passing we remark that our approach is
instructive in yet another respect, namely that our smearing distribution
$\omega(v,a,t)$ is inevitably time-dependent. Though superstatistics
does not prohibit {\em per se} time-dependent smearing
distributions, common practice is to assume (at least in first
approximation) that the fluctuation parameter (inverse temperature,
volatility, energy dissipation rate, etc.) as well as its moments  do not have
explicit time dependence. This is not the case here since
$\langle v^\alpha \rangle \propto t^{\alpha}$
for $\alpha < 1/2$. In our considerations we have two well-separated time
scales: a short time scale of order $\varDelta t$ which corresponds
to the size of the time mesh, and  a macroscopic time $t$ ($t\gg
\varDelta t$) over which $\omega$ changes significantly ---
$t$ is proportional to a statistical dispersion (scale parameter) of
$\omega$ (see, e.g., \cite{Weibull51}). An explicit use of the
macroscopic time $t$ in $\omega$ is mandatory and we should not ignore
it if we want to obtain correct relativistic propagators. Note also
that because $\langle v^\alpha \rangle$ diverges for  $\alpha > 1/2$,
one cannot apply any form of truncated cumulant expansion (often used
in perturbative superstatistics) to obtain, e.g., a
non-relativistic limit. The path integral identity
(\ref{SEc4.1a}) is fully non-perturbative in $v$.

In the end we wish to add few more comments concerning worldline
path integrals. Worldline representations of field-theoretic
propagators, as considered here, are an aspect of the so-called
``worldline quantization" of particle physics. In this approach the
process of second quantization is reversed. Second quantization, or
field quantization, was introduced to represent a grand canonical
ensemble of quantum particles by a single quantum field. Since each
quantum particle possesses a fluctuating world line, quantized field
theory is the most efficient way of studying grand-canonical
ensembles of fluctuating lines. These can be, for instance,
worldlines of elementary particles as emphasized by
Feynman~\cite{Feynman50} and Schwinger~\cite{Schwinger:51}, or lines
of a completely different physical nature, such as polymers,
vortices or defect lines. In the latter case it is possible to study
the phase transitions caused by the proliferation of such vortices
or defect lines with the help of a single quantum field. The
associated quantum field theory is known  as {\em disorder field
theory\/}~\cite{kleinert:89}. In the first case, the phase
transitions in polymer ensembles become tractable by the efficient
methods of quantum field theory~\cite{PI}.
At a phase transition, an order or a disorder field
can acquire a nonzero expectation value. This phenomenon is very hard
to describe in a first-quantized world-line approach \cite{BEC}.

In many recent works, this development has been turned around. The
motivation for this comes from the inability to develop a
second-quantized field theory for strings, whose ``worldlines" are
fluctuating surfaces (worldsheets). In string theory, calculations
have so far remained restricted to the first-quantized
formulation~\cite{excep}. In order to gain more insight
people~\cite{polyakov:87,Schubert} have returned to well-understood
quantum field theoretic problems of point particles and reconsidered
them in the first-quantized formulation in which fluctuating
worldlines play the essential role. This, so-called
``string-inspired" approach has led to a great number of
publications initiated by Bern and Kosewer~\cite{Bern:91,Schubert}.
They shed an alternative light on calculations within quantum
electrodynamics (QED)~\cite{affleck:82} and quantum chromodynamics
(QCD)~\cite{Schubert,antonov:00}, on calculations of
anomalies~\cite{Bastinielli:92} and of index densities in the
Atiyah-Singer theorem~\cite{alvarez:83}. Besides, worldline
quantization forms an integral part of the so-called operator
regularization scheme of McKeon {\em et al.}~\cite{mckeon}.



\appendix
\section{\label{ap1a}}

Here we briefly discuss the Lorentz properties
 of the two-component wave function
\begin{equation}
\Psi({\bi x},t) =
\left(
                                                                  \begin{array}{c}
 \phi({\bi x},t) \\
                                                                     \chi({\bi x},t) \\
                                                                   \end{array}
                                                                 \right).
\label{A.1abc}
\end{equation}
We first observe that the components $\phi$ and $\chi$ can be represented as
\begin{eqnarray}
\phi \ = \ \frac{1}{\sqrt{2}} \left(\psi \ - \ \frac{1}{\rmi m c^2} \frac{\partial \psi }{\partial t} \right),\nonumber \\[2mm]
\chi \ = \ \frac{1}{\sqrt{2}} \left(\psi \ + \ \frac{1}{\rmi m c^2} \frac{\partial \psi}{\partial t} \right),
\label{A.1abcd}
\end{eqnarray}
where $\psi$ is a Klein-Gordon field fulfilling
\begin{eqnarray}
(\square \ + \  m^2c^2)\psi \ = \ 0\, .
\label{A.1abce}
\end{eqnarray}
Combining (\ref{A.1abcd}) with (\ref{A.1abce}) one can easily
check that $\Psi$  satisfies the Schr\"{o}dinger
equation (\ref{2.1}) with the Hamiltonian $H_{\rm FV}$ given by (\ref{2.2}).
Using the fact that $\psi$ is a Lorenzian scalar one can deduce the transformation properties of $\Psi$ under the Lorentz group as follows:
Under finite Lorentz transformation $\Lambda$ the field $\Psi$ should transform as
\begin{eqnarray}
\Psi(x) \ \stackrel{\Lambda}{\longrightarrow} \  \Psi'(x) \ = \ S({\Lambda})\Psi(\Lambda^{-1}x)\, ,
\label{A.1abcaa}
\end{eqnarray}
where $S({\Lambda})$ represents an operator of intrinsic field transformations.
Eq.~(\ref{A.1abcaa}) implies an infinitesimal Lorentz transformation
\begin{eqnarray}
\Psi(x) \ \stackrel{\Lambda}{\longrightarrow} \  \Psi(x) \ + \ \delta_{\Lambda}\! \Psi(x)\, .
\end{eqnarray}
Here
\begin{eqnarray}
\delta_{\Lambda}\! \Psi(x) \ = \ \Psi'(x)  -  \Psi(x) \,\,\,\,\, \mbox{and}  \,\,\,\,\, \delta_{\Lambda} x^{\mu} \ = \ x'^{\mu}  -  x^{\mu} \ = \ \omega_{\;\;\nu}^{\mu} x^{\nu} \ = \  - \frac{\rmi}{2} \omega_{\mu\nu}(S^{\mu \nu})x
 \, .
\end{eqnarray}
where the antisymmetric matrix $\omega_{\mu\nu} = - \omega_{\nu\mu}$
collects both rotation angles and rapidities, i.e., $\omega_{ij} =
\epsilon_{ijk}\varphi^k$ and $\omega_{0i} = \zeta^i = p^i/mc$, respectively.
$(S^{\mu\nu})_{\alpha}^{\;\;\beta} = \rmi (\eta_{\mu\alpha}\eta^{\beta}_{\;\;\nu} - \eta_{\nu\alpha}\eta^{\beta}_{\;\;\mu})$ represent generators of the Lorentz group for  vectors.

We may now employ the fact that $\delta_{\Lambda}  \psi =
-\frac{i}{2}
\omega_{\mu \nu} \hat{L}^{\mu \nu} \psi$, where
$\hat{L}_{\mu \nu} = \rmi(x_{\mu}\partial_{\nu}
- x_{\nu} \partial_{\mu})$ represent the
generators of the Lorentz group for scalar fields, and write
\begin{eqnarray}
\delta_{\Lambda}\! \Psi(x) \ = \ \frac{1}{\sqrt{2}}\left(
                                                     \begin{array}{c}
                                                       -\frac{i}{2} \omega_{\mu \nu} \hat{L}^{\mu \nu} \psi \ - \ \frac{1}{\rmi mc}\ \! \delta_{\Lambda} \partial_0 \psi \\[2mm]
                                                       -\frac{i}{2} \omega_{\mu \nu} \hat{L}^{\mu \nu} \psi \ + \ \frac{1}{\rmi mc} \ \! \delta_{\Lambda} \partial_0 \psi \\
                                                     \end{array}
                                                   \right).
                                                   \label{A.5abc}
\end{eqnarray}
If we now utilize the property
\begin{eqnarray}
\delta_{\Lambda} \partial_0 \psi \ = \ -\frac{\rmi}{2}\ \! \omega_{\mu \nu}\hat{L}^{\mu \nu} \ \!\partial_0 \psi \ - \ \frac{\rmi}{2}\ \!(\omega_{\mu \nu} S^{\mu \nu})^{\;\; \alpha}_0 \ \! \partial_{\alpha} \psi\, ,
\end{eqnarray}
we can cast (\ref{A.5abc}) into a form
\begin{eqnarray}
\delta_{\Lambda}\! \Psi(x) \ = \  -\frac{i}{2} \omega_{\mu \nu} \left[ \hat{L}^{\mu \nu} \ + \ \frac{1}{2mc} (S^{\mu\nu})_0^{\;\;\alpha} \hat{p}_{\alpha} (\sigma_3 + \rmi \sigma_2)\right]\Psi(x)\, ,
\label{A.6abc}
\end{eqnarray}
which identifies the generators of the
Lorentz transformations on the two-component wave  Feshbach-Villars wave function $\Psi(x)$  as
\begin{eqnarray}
\hat{M}^{\mu\nu} \ = \ \hat{L}^{\mu \nu} \ + \ \frac{1}{2mc} \ \! (S^{\mu\nu})_0^{\;\;\alpha} \hat{p}_{\alpha} (\sigma_3 + \rmi \sigma_2)\, .
\label{A.7abc}
\end{eqnarray}
In particular, if $\Lambda$ describes rotations,
then  $\omega_{\mu\nu}$ has only spatial indices and
$(S^{\mu\nu})_0^{\;\;\alpha} \mapsto (S^{ij})_0^{\;\;\alpha} =
0$. This implies that $\hat{M}_{ij} = \hat{L}_{ij}= \rmi(x_i\partial_j
- x_j \partial_i)$, which are standard generators of rotation for scalar fields.
If $\Lambda$ corresponds to  boost transformations,
then $(S^{\mu\nu})_0^{\;\;\alpha} \mapsto (S^{0j})_0^{\;\;i} \neq 0$, and the boost generators read
\begin{eqnarray}
\hat{K}_i \ \equiv \ \hat{M}^{0i} \ = \ \hat{L}^{0i} + \frac{1}{2mc}\ \!
S^{0i}\hat{\bi p}(\sigma_3 + \rmi \sigma_2)\, .
\label{A.7abcf}
\end{eqnarray}
Here the product $S^{0i}\hat{\bi p}$ is defined by the contraction
$(S^{0i})_0^{\;\;k}\hat{p}_k$.

Let us now show that the generators $\hat{M}^{\mu\nu}$ close the
$SO(3,1)$ algebra. As for generators $\hat{M}_{ij} $, these clearly constitute the rotational
sub-algebra $SO(3)$, i.e.,
\begin{eqnarray}
[\hat{J}_i, \hat{J}_j] \ = \ \rmi \epsilon_{ijk} \hat{J}_k\, \;\;\;\;\;
\mbox{with}\;\;\;\;\; \hat{J}_i \ = \ \frac{1}{2}
\epsilon_{ijk}\hat{M}^{jk}\, . \label{A.2abc}
\end{eqnarray}
The generators $\hat{K}_i \equiv \hat{M}^{0i}$ yield commutators
\begin{eqnarray}
[\hat{K}_i, \hat{K}_j] \ &=& \ [\hat{L}^{0i} + \frac{1}{2mc}\ \!
S^{0i}\hat{\bi p}(\sigma_3 + \rmi \sigma_2), \hat{L}^{0j} + \frac{1}{2mc} \ \!
S^{0j}\hat{\bi p}(\sigma_3 + \rmi \sigma_2) ]\nonumber \\[2mm]
&=& \ [\hat{L}^{0i}, \hat{L}^{0j}] \ + \ \frac{1}{2mc}\ \!
(S^{0i})^0_{\;\;k}\ \![\hat{p}^k, \hat{L}^{0j}](\sigma_3 + \rmi
\sigma_2) \ + \ \frac{1}{2mc} \ \!(S^{0j})^0_{\;\;k}\ \![\hat{L}^{0i},
\hat{p}^k](\sigma_3 + \rmi \sigma_2)\nonumber \\[2mm]
&=& \ [\hat{L}^{0i}, \hat{L}^{0j}] \ + \ \frac{i}{2mc} \ \!
(S^{0i})^0_{\;\;k}\ \! \hat{p}^0 \delta^{kj}(\sigma_3 + \rmi \sigma_2)
\ - \ \frac{i}{2mc} \ \! (S^{0j})^0_{\;\;k}\ \! \hat{p}^0 \delta^{ik}
(\sigma_3 + \rmi \sigma_2)
\nonumber \\[2mm]
&=& \ [\hat{L}^{0i}, \hat{L}^{0j}] \ = \ -\rmi \epsilon_{ijk}
\hat{J}_k\, . \label{A.3abc}
\end{eqnarray}
Which is a familiar boost commutator. In the derivation we have used
the fact that $(\sigma_3 + \rmi \sigma_2)^2 = 0$ and  that
$\hat{L}^{0i}$ are boost generators for scalar fields.
Finally, the mixed commutators
read
\begin{eqnarray}
[\hat{J}_i, \hat{K}_j] \ &=& \ [\hat{J}_i, \hat{L}^{0j} +
\frac{1}{2mc}
S^{0j}\hat{\bi p}(\sigma_3 + \rmi \sigma_2)]\nonumber \\[2mm]
&=& \ [\hat{J}_i, \hat{L}^{0j}] \ + \ \frac{1}{2mc}(S^{0j})^{0k}\
\![\hat{J}_i, \hat{p}_k](\sigma_3 + \rmi \sigma_2)\nonumber \\[2mm]
&=& \
\rmi\epsilon_{ijk} \hat{L}^{0k} \ - \ \frac{\rmi}{2mc}(S^{0j})^{0k}\
\!\epsilon_{ikl}
\hat{p}^{l} (\sigma_3 + i \sigma_2)\nonumber \\[2mm]\
&=& \ \rmi \epsilon_{ijk} \left(\hat{L}^{0k} \ + \ \frac{1}{2mc}
S^{0k}\hat{\bi p}(\sigma_3 + \rmi \sigma_2) \right) \ = \ \rmi
\epsilon_{ijk} \hat{K}_k .\label{A.4abc}
\end{eqnarray}
Here we have utilized the identity $(S^{0j})^{0k}
\epsilon_{ikl}\hat{p}^l = \rmi\eta^{jk} \epsilon_{ikl}\hat{p}^l = -
\rmi\epsilon_{ijl}\hat{p}^l = - (S^{0k})^{0}_{\;\; l}
\epsilon_{ijk}\hat{p}^l$. As a result we see the commutators
(\ref{A.2abc}), (\ref{A.3abc}), and (\ref{A.4abc}) close the Lorentz
algebra $SO(3,1)$.

\section{\label{ap1}}

In this appendix we show that ${U}_{{\bi p}}$ may be viewed as a
boost transformation that brings a wave function $ \Psi({\bi x},t)$
of a spinless particle at rest to the velocity ${\bi v}$. To this
end we seek the positive- and negative-energy plane wave
solutions of the Schr\"{o}dinger-like equation (\ref{2.1}) in the
form
 \begin{eqnarray}
&&\Psi^{(+)}({\bi x},t)\ = \  \! u(p)\,\rme^{-\rmi px}\ ,\nonumber \\
&&\Psi^{(-)}({\bi x},t)\ = \   \! v(p)~\rme^{\rmi px}\, , \label{A1a}
\end{eqnarray}
with
\begin{eqnarray}
&&(cp_0 \ - \ {H}_{{\bi p}} {U}_{{\bi p}}  \sigma_3 {U}_{{\bi p}}^{-1}) u(p)
\ = \ 0\, , \nonumber \\
&&(cp_0 \ + \ {H}_{{\bi p}}{U}_{{\bi p}} \sigma_3 {U}_{{\bi
p}}^{-1}) v(p) \ = \ 0\, , \label{A2a}
\end{eqnarray}
where
$p_0 = \sqrt{{\bi p}^2+m^2c^2} $. For
the rest momentum $p_R \equiv (m c, {\bi
0})$, these equations simplify
 and the respective amplitudes $u(p_R)$ and $v(p_R)$ satisfy
\begin{eqnarray}
&&( \sigma _0 - \sigma_3)u(p_R) \ = \ 0\, ,\nonumber \\
&& ( \sigma _0 + \sigma_3)v(p_R) \ = \ 0 \, .
\end{eqnarray}
The solutions are
\begin{eqnarray}
u(p_R) \ = \ \left( \begin{array}{c}
                              1 \\
                              0
                         \end{array} \right)\! ,
 \;\;\;\;\;\;\;\;\;\; v(p_R) \ = \ \left( \begin{array}{c}
                              0 \\
                              1
                         \end{array} \right)\! .
\end{eqnarray}
They are normalized to ensure the unit normalization (\ref{21a}):
\begin{eqnarray}
(u,u) \ = \ u^\dag\sigma_3 u \ = \ 1\, , \;\;\; (v,v) \ = \
v^\dag\sigma_3 v \ = \ - 1\, , \;\;\;  (u,v) \ = \ u^\dag\sigma_3 v
\ = \ 0\, ,
\label{A.5}
\end{eqnarray}
making positive and negative energy states orthogonal to each other. Using the identity
\begin{eqnarray}
(cp_0 \ \pm \ {H}_{{\bi p}}{U}_{{\bi p}} \sigma_3  {U}_{{\bi p}}^{-1})
(cp_0 \ \mp \ {H}_{{\bi p}}{U}_{{\bi p}}\sigma_3  {U}_{{\bi p}}^{-1})
\ = \ c^2 p_0^2 \ - \ {H}_{{\bi p}}^2 \ = \  0\, ,
\end{eqnarray}
we can write the amplitudes $u(p)$ and $v(p)$ at arbitrary momentum
(cf. Eqs.~(\ref{A2a})) as
\begin{eqnarray}
u(p)  \ = \ N_p  (cp_0 + {H}_{{\bi p}}{U}_{{\bi p}}\sigma_3  {U}_{{\bi
p}}^{-1})\left(
\begin{array}{c}
                             1  \\
                              0
                         \end{array} \right) \ = \ N_p\left( \begin{array}{c}
                              cp_0 + {\bi p}^2/2m + mc^2 \\
                              -{\bi p}^2/2m
                         \end{array} \right)\! ,
\label{A7}
\end{eqnarray}
and
\begin{eqnarray}
v(p) \ = \ N_p (cp_0 - {H}_{{\bi p}}{U}_{{\bi p}}\sigma _3 {U}_{{\bi
p}}^{-1})\left( \begin{array}{c}
                              0 \\
                             1                         \end{array} \right) \ = \ N_p\left( \begin{array}{c}
                               -{\bi p}^2/2m \\
                               cp_0 + {\bi p}^2/2m + mc^2
                         \end{array} \right)\! ,
\label{A8}
\end{eqnarray}
with some normalization constant $N_p$.
The normalization conditions (\ref{A.5}) require
\begin{eqnarray}
N_p \ = \ \sqrt{\frac{mc}{p_0}} \ \!\frac{1}{cp_0 + mc^2}\, ,
\end{eqnarray}
so that Eqs.~(\ref{A7}) and~(\ref{A8}) become
\begin{eqnarray}
&&u(p)  \ = \ \frac{1}{2\sqrt{mc p_0}}\left( \begin{array}{c}
                          {mc + p_0}\\
                          {mc - p_0}
\end{array} \right),~~~~
v(p)  \ = \ \frac{1}{2\sqrt{mc p_0}}\left( \begin{array}{c}
                          {mc - p_0}\\
                          {mc + p_0}
\end{array} \right).
\label{A11}
\end{eqnarray}
Eqs.~(\ref{A11}) define boost transformations
%
%
 \begin{eqnarray}
&&u(p)  \ = \
U_{\bi p}
\left(
\begin{array}{c}
              1\\
              0
\end{array} \right)
,~~~
v(p)  \ = \
U_{\bi p}
\left(
\begin{array}{c}
              0\\
              1
\end{array} \right)
.
\label{A12w}
\end{eqnarray}
Here we have denoted the boost matrix as $U_{\bi p}$
because it appears here in  the form
\begin{equation}
 U_{\bi p}\ = \
  \frac{p_0 + mc}{2\sqrt{mc p_0}}\ \!\sigma_0
\ - \   \frac{p_0- mc}{2\sqrt{mc p_0}}\ \! \sigma_1 \, ,
\label{MB}\end{equation}
which is identical to the diagonalization matrix $ U_{\bi p}$ as defined
by Eq.~(\ref{2.3a}). If we introduce a parameter
\begin{equation}
\alpha_{\bi v}  \ \equiv \   \ln \sqrt{\frac{p_0}{mc}} \ = \ \frac{1}{2}
\ln  \gamma _{\bi v}\, ,
\label{A22a}\end{equation}
which satisfies
\begin{eqnarray}
\cosh \alpha _{\bi v}  \ = \ \frac{p_0 + mc}{2\sqrt{mc p_0}} \;\;\;\;\;\;
\mbox{and} \;\;\;\;\;\; \sinh \alpha _{\bi v}  \ = \ \frac{
p_0 - mc}{2\sqrt{mc p_0}}\, ,
\end{eqnarray}
then we can write the boost matrix (\ref{MB}) as an exponential (cf. also Eq.~(\ref{2.3a}))
\begin{eqnarray}
U _{\bi p}
 = \ \exp\left( -  \alpha _{\bi v}
\sigma_1 \right).
 \label{a13a}
\end{eqnarray}
Connection of $U _{\bi p}$ with boost generators (\ref{A.7abcf}) can be established when we rewrite (\ref{A.6abc}) for boost transformation in the form
\begin{eqnarray}
\Psi'(x') \ = \  \left(1  \ - \ \frac{\rmi}{2mc} \ \! \zeta^i \ \! (S^{0 i}) \hat{\bi p} \ \! (\sigma_3 + \rmi \sigma_2)\right)\Psi(x)\, .
\label{a14a}
\end{eqnarray}
For positive-energy plane wave solutions this can be written as
\begin{eqnarray}
u(p') \ &=& \ \left(1  \ - \ \frac{\rmi}{2mc} \ \! \zeta^i \ \! (S^{0 i}){\bi p} \ \!(\sigma_3 + \rmi \sigma_2)\right) u(p)\nonumber \\[1mm]
&=& \ \frac{1}{\sqrt{2}}\left(
                                                 \begin{array}{c}
                                                  1 \ + \ p^0/mc \ + \ \zeta^i p^i/mc \\
                                                  1 \ - \ p^0/mc \ - \ \zeta^i p^i/mc  \\
                                                 \end{array}
                                               \right)
 \tilde{\psi}^{+}(p)\, .
\label{a15a}
\end{eqnarray}
Here, $\tilde{\psi}^{+}(p)$ stands for an amplitude of the positive-energy plane wave solution of the Klein-Gordon equation. Term $p^0  +  \zeta^i p^i$ can be recognized as a first-order term in the Lorentz boost transformation
\begin{eqnarray}
\Lambda(\bi \zeta)^0_{\,\,\mu}\ \! p^{\mu} \ = \ p^0 \ \! \cosh \zeta   \  +  \ (\hat{\bi \zeta}\cdot {\bi p}) \ \! \sinh \zeta \ = \ p'^0\, .
\end{eqnarray}
Here $\hat{\bi \zeta} = {\bi u}/|{\bi u}|$ denotes the unit vector in the direction of the boost velocity ${\bi u}$ (${\bi u}\oplus {\bi v} = {\bi v}'$). If we further employ the identities:
\begin{eqnarray}
\cosh \zeta \ = \  \gamma_{{\bi u}}\;\;\;\;\;\;\;\; \mbox{and} \;\;\;\;\;\;\;\; \hat{\bi\zeta} \ \! \sinh \zeta \ = \ \gamma_{{\bi u}} \frac{{\bi u}}{c}\, ,
\end{eqnarray}
we may cast (\ref{a15a}) into form
\begin{eqnarray}
\mbox{\hspace{-1cm}}u(p') \ &=& \ \frac{1}{\sqrt{2}}\left(
                                                 \begin{array}{c}
                                                  1 \ + \ \gamma_{{\bi u}}\gamma_{{\bi v}}(1 + {\bi u}\cdot {\bi v}/c^2) \\
                                                  1 \ - \ \gamma_{{\bi u}}\gamma_{{\bi v}}(1 + {\bi u}\cdot {\bi v}/c^2) \\
                                                 \end{array}
                                               \right)
 \tilde{\psi}^{+}(p) \ = \ \frac{1}{\sqrt{2}}\left(
                                                 \begin{array}{c}
                                                  1 \ + \ \gamma_{{\bi u}\oplus {\bi v}} \\
                                                  1 \ - \ \gamma_{{\bi u}\oplus {\bi v}} \\
                                                 \end{array}
                                               \right)
 \tilde{\psi}^{+}(p)\, .
\label{a15accc}
\end{eqnarray}
Which clearly shows that the original amplitude was boosted from the velocity ${\bi v}$ to the amplitude with the velocity ${\bi v}' = {\bi u}\oplus{\bi v}$.

In the particular case when the initial momentum is $p_r$, the relation (\ref{a15accc})
acquires the form
\begin{eqnarray}
u(p) \ &=&
\frac{1}{\sqrt{2}}\left(
                                                 \begin{array}{c}
                                                  1 \ + \ \gamma_{{\bi u}} \\
                                                  1 \ - \ \gamma_{{\bi u}} \\
                                                 \end{array}
                                               \right)
 \tilde{\psi}^{+}(p_r)\, .
\label{a15acccdd}
\end{eqnarray}
By utilizing the normalization condition (\ref{A.5}) we have that $\tilde{\psi}^{+}(p_r) = 1/\sqrt{2 \gamma_{{\bi u}}}$, which finally gives
\begin{eqnarray}
u(p) \ =  \
\frac{1}{{2}}\left(
                                                 \begin{array}{c}
                                                  1/\sqrt{\gamma_{{\bi u}}} \ + \ \sqrt{\gamma_{{\bi u}}} \\
                                                  1/\sqrt{\gamma_{{\bi u}}} \ - \ \sqrt{\gamma_{{\bi u}}} \\
                                                 \end{array}
                                               \right) \ = \  U_{{\bi u}} u(p_r)\, .
\label{a16acccdd}
\end{eqnarray}
Analogous analysis applies for negative-energy plane waves in which case we obtain
\begin{eqnarray}
v(p) \ =  \
\frac{1}{{2}}\left(
                                                 \begin{array}{c}
                                                  1/\sqrt{\gamma_{{\bi u}}} \ - \ \sqrt{\gamma_{{\bi u}}} \\
                                                  1/\sqrt{\gamma_{{\bi u}}} \ + \ \sqrt{\gamma_{{\bi u}}} \\
                                                 \end{array}
                                               \right) \ = \ U_{{\bi u}} v(p_r)\, .
\label{a17acccdd}
\end{eqnarray}
Results (\ref{a16acccdd}) and (\ref{a17acccdd}) establish the promised identification between $U_{{\bi u}}$ and boost transformations from the rest frame velocity ${\bi v} = {\bi 0}$ to the velocity ${\bi u}$.

Let us finally comment on  a non-relativistic limit of the Feshbach-Villars wave function.  To this end we approximate
$H_{\bi p} = c\sqrt{{\bi p}^2 + m^2c^2}\approx mc^2  + {\bi p}^2/2m$.
With this the positive- and negative-energy solution (\ref{A1a}) become
\begin{eqnarray}
&&\Psi^{(+)}({\bi x}, t) \ \stackrel{c\rightarrow \infty}{\approx} \ \hspace{-1pt} \left( \begin{array}{c}
                              1 \\
                              -{\bi v}^2/4c^2
                         \end{array} \right)\exp[\rmi({\bi p}\cdot {\bi x} - H_{\bi p}t)] \ \equiv \
\Phi^{(+)}({\bi x}, t)\ \! e^{-\rmi mc^2 t}\, ,\nonumber \\[2mm]
&&\Psi^{(-)}({\bi x}, t) \ \stackrel{c\rightarrow \infty}{\approx} \  \left( \begin{array}{c}
                               -{\bi v}^2/4c^2\\
                              1
                         \end{array} \right)\exp[\rmi( H_{\bi p}t -{\bi p}\cdot {\bi x})]\ \equiv \
\Phi^{(-)}({\bi x}, t)\ \! e^{\rmi mc^2 t}\, .
\end{eqnarray}
In particular, we see that for plane particle waves, the upper components
are much larger than the lower components.
The opposite holds for antiparticle waves.
This analogous to the situation for Dirac wave function. For particle waves,
Eqs.~(\ref{II.13a0}) and (\ref{II.13a}) reduce  to
\begin{eqnarray}
&&\rmi \partial_t \Phi^{(+)}  \ = \ -\frac{\nabla^2}{2m} \sigma_3 \ \!\Phi^{(+)}\,\label{A16a},~~~~
\rmi \partial_t \Phi^{(-)}  \ = \ -\frac{\nabla^2}{2m} \sigma_3 \ \!\Phi^{(-)}\, .\label{A17a}
\end{eqnarray}
By neglecting the small component in $\Phi^{(+)}$,
this implies the Schr\"{o}dinger equation for the large component in  $\Phi^{(+)}$
with $\hat H = -\nabla^2/2m$. An analogous situation holds for $\Phi^{(-)}$.

%

\section*{Acknowledgments}

This work was partially supported by the Ministry of
Education of the Czech Republic (research plan MSM 6840770039),
and by the Deutsche Forschungsgemeinschaft under grant Kl256/47.



\section*{References}
\begin{thebibliography}{10}
\bibitem{beck:01} C.~Beck, {\em Phys.\ Rev.\ Lett.} {\bf 87} (2001) 180601
\bibitem{beck:03} C.~Beck and E.G.D.~Cohen, {\em Physica} {\em A~}{\bf 322} (2003) 267
\bibitem{wilk:00} G.~Wilk and Z.~Wlodarczyk, {\em Phys.\ Rev.\ Lett.} {\bf 84} (2000) 2770
\bibitem{touchette:05} H.~Touchette and C.~Beck, {\em Phys.\ Rev.  E~}{\bf 71} (2005) 016131
\bibitem{sattin:04} F.~Sattin, {\em Physica} {\em A~}{\bf338} (2004) 437
\bibitem{beck:05} C.~Beck, E.G.D.~Cohen and H.L.~Swinney, {\em Phys.\ Rev. E~}{\bf 72} (2005) 56133
\bibitem{vignat:05} C.~Vignat and A.~Plastino, [arXiv:0706.0151]
\bibitem{chavanis:06} P.~H.~Chavanis, {\em Physica} {\em A~}{\bf 359} (2006) 177
\bibitem{JK:08} P.~Jizba and H.~Kleinert, {\em Phys.\ Rev.  E~}{\bf 78} (2008)
031122
\bibitem{jkh07} P.~Jizba, H.~Kleinert and P.~Haener, {\em Physica} {\em A~}(2009) 3503
\bibitem{beck:09} C.~Beck, [arXiv:0811.4363]

\bibitem{Dunning-Davies:05} R.~Kubo, M.~Toda and N.~Hashitsume,
{\em Statistical Physics II: Nonequilibrium Statistical Mechanics}
(Springer, New York, 1995)); see also, J.~Dunning-Davies,
[arXiv:physics/0502153]
\bibitem{lavenda:91} B.H.~Lavenda, {\em Statistical Physics: A Probabilistic
Approach} (Wiley-Interscience, New York, 1991)
\bibitem{feller66} W.~Feller, {\em An Introduction to Probability Theory and its Applications, Vol. II}
(John Wiley, London, 1966)

\bibitem{Weibull51} W.~Weibull, {\em J. Appl. Mech.-Trans. ASME} {\bf 18} (1951) 293

\bibitem{PI} H.~Kleinert, {\em Path Integrals in Quantum Mechanics, Statistics,
Polymer Physics and Financial Markets} (World Scientific, Singapore
2009) (http://www.physik.fu-berlin.de/\~{}kleinert/b5)

\bibitem{Sorensen:02} see e.g., D.~Sorensen and D.~Gianola, {\em Likelihood, Bayesian, and
MCMC methods in quantitative genetics} (Springer-Verlag,
New York, 2002)
\bibitem{hartle:01} J.B.~Hartle and K.V.~Kucha\v{r},
{\em Phys.\ Rev. D~}{\bf 34} (1986) 2323
\bibitem{Grosche:98} C.~Grosche and F.~Steiner, {\em Handbook of Feynman Path
Integrals} (Springer, Berlin, 1998)
\bibitem{polyakov:87} A.M.~Polyakov, {\em Gauge Fields and Strings} (Harwood,
New York, 1987)
\bibitem{Gitman:90} D.M.~Gitman and I.V.~Tyutin,
{\em Quantization of Fields with Constraints},
(Springer-Verlag, New York, 1990)
\bibitem{Sundermeyer:1982gv}
  K.~Sundermeyer,
  {\em Constrained Dynamics With Applications To Yang-Mills Theory, General
  Relativity, Classical Spin, Dual String Model}, (Springer-Verlag, Berlin, 1982)
\bibitem{Feshbach58} H.~Feshbach and F.~Villars, {\em Rev. \ Mod. \ Phys} {\bf
30} (1958) 24
\bibitem{Feynman:49} R.P.~Feynman, {\em Phys.\ Rev.} {\bf 76} (1949) 749
\bibitem{Stuckelberg:41} E.C.G.~St\"{u}ckelberg, {\em Helvetica Physica Acta} {\bf 14} (1941) 588
\bibitem{Stuckelberg:42} E.C.G.~St\"{u}ckelberg, {\em Helvetica Physica Acta} {\bf 15} (1942) 23
\bibitem{Feynman50} R.P.~Feynman, {\em Phys.\ Rev.} {\bf 80} (1950) 440
\bibitem{Foldy50} L.L.~Foldy and S.A.~Wouthuysen, {\em Phys.\ Rev.} {\bf 78} (1950) 29
\bibitem{zitt}
Zitterbewegung is a rapid oscillatory motion of a relativist
particle caused by interference between positive and negative energy
parts of the wave function.
\bibitem{Z-J:04} J.~Zinn-Justin, {\em Quantum Field Theory and Critical Phenomena}, (Clarendon Press, Oxford, 2002)
\bibitem{Polchinski} J.~Polchinski, {\em String Theory, Vol. 1}, (Cambridge University Press, Cambridge, 2005)
\bibitem{case:54} K.M.~Case, {\em Phys. Rev.} {\bf 95} (1954) 1323
\bibitem{jorgensen} B.~J{\o}rgensen, {\em Statistical Properties of the Generalized Inverse Gaussian
Distribution}, Lecture Notes in Statistics 9, (Springer-Verlag,
Berlin, 1982)
\bibitem{bohm:96} D.~Bohm, {\em The Special Theory of Relativity}, (Routledge, New York, 1996)
\bibitem{bohm:93} D.~Bohm and B.J.~Hiley, {\em The Undivided Universe}, (Routledge, New York, 1993)
\bibitem{garay:98} L.J.~Garay, {\em Phys. Rev. Lett.} {\bf 80} (1998) 2508
\bibitem{Schwinger:51} J.~Schwinger, {\em Phys.\ Rev.} {\bf 82} (1951) 664
\bibitem{gambini:99} R.~Gambini and J.~Pullin {\em J. Phys. Rev. D~}{\bf 59} (1999) 124021
\bibitem{alfaro:00} J.~Alfaro, H.A.~Morales-Tecotl and L.F.~Urrutia {\em Phys. Rev. Lett.} {\bf 84} (2000) 2318
\bibitem{camelia:00} G.~Amelino-Camelia and S.~Majid  {\em Int. J. Mod. Phys. A~}{\bf 15} (2000) 4301
\bibitem{susskind:00} A.~Matusis, L.~Susskind and N.~Toumbas {\em J. High Energy Phys.} JHEP0012 (2000) 002
\bibitem{camelia:01} G.~Amelino-Camelia  {\em Phys. Lett. B}~{\bf 510} (2001) 255
\bibitem{douglas:01} N.R.~Douglas and N.A.~Nekrasov  {\em Rev. Mod. Phys.} {\bf 73} (2001) 977
\bibitem{jacobson:08} T.~Jacobson and A.C.~Wall, [arXiv:0804.2720]
\bibitem{eriksen:58} E.~Eriksen, {\em Phys. Rev.~}{\bf 111} (1958) 1011
\bibitem{kleinert:89} H.~Kleinert, {\em Gauge fields in condensed matter. Vol. 1: Superflow and
vortex lines. Disorder fields, phase transitions} (World Scientific,
Singapore, 1989); (http://www.phy\-sik.fu-ber\-lin.de/\~{}klei\-nert/b1)
\bibitem{BEC}
See the treatment of Bose-Einstein
condensation in Chapter 7 in the textbook \cite{PI}.
\bibitem{excep}
For an early proposal of a second-quantized field
of ensembles of strings see: H.~Kleinert, {\em Lettere Nuovo Cimento}   {\bf 4} (1970) 285;
(http://www.phy\-sik.fu-ber\-lin.de/\~{}klei\-nert/24)
\bibitem{Schubert} C.~Schubert, {\em Physics Report} {\bf 355} (2001) 73
\bibitem{Bern:91} Z.~Bern and D.A.~Kosower,  {\em Phys.\ Rev.\ Lett.} {\bf 66} (1991)
1669; Z.~Bern and D.A.~Kosower, {\em Nucl. Phys.}  {\em B~}{\bf 379}
(1992) 451
\bibitem{affleck:82} I.K.~Affleck, O.~Alvarez and N.S.~Manton,  {\em Nucl.
Phys.}  {\em B~}{\bf 197} (1982) 509
\bibitem{antonov:00} D.~Antonov, {\em Phys. Lett.} {\em B~}{\bf 479} (2000)
387
\bibitem{Bastinielli:92} L.~Alvarez-Gaum\'{e} and E.~Witten, {\em Nucl.
Phys.} {\em B~}{\bf 234} (1983) 269; F.~Bastinielli and
P.~van~Niewenhuizen, {\em Nucl. Phys.}  {\em B~}{\bf 389} (1993) 53
\bibitem{alvarez:83} L.~Alvarez-Gaum\'{e}, {\em Comm. Math. Phys.}
{\bf 90} (1983) 161;  D.~Friedan and P.~Windey, {\em Nucl. Phys.}
{\em B~}{\bf 234} [FS11] (1984) 395
\bibitem{mckeon} D.G.C.~McKeon and T.N.~Sherry, {\em Phys.\ Rev.}
{\bf 35}  (1987) 3854; D.G.C.~McKeon and C.~Wong, {\em J. Math. Phys.} {\bf 36} (1995) 1691.
\bibitem{wigner:39} E.P.~Wigner, {\em Ann. Math.~}{\bf 40} (1939)
149; Wigner's (or also Thomas-Wigner's) rotation is a
group-theoretic consequence of the algebraic structure of the
Lorentz group --- algebra of boosts is not closed, one needs extra
rotation to close the algebra.
\bibitem{ferraro:99} R.~Ferraro and M.~Thibeault, {\em Eur. J.
Phys.~}{\bf 20} (1999) 143
\bibitem{Simon:1990} R.~Simon and N.~Makunda, {\em Found. Phys. Let.~}{\bf 3} (1990) 425
\bibitem{perelomov} A.~Perelomov, {\em Generalized Coherent States and Their Applications} (Springer-Verlag, Berlin, 1986)
%
%




\end{thebibliography}
\end{document}